\documentclass[aps, pra, showkeys, reprint, nofootinbib, longbibliography]{revtex4-1} 

\usepackage[usenames,dvipsnames]{color} 
\usepackage{amsmath} 
\usepackage{txfonts} 
\usepackage{upgreek} 
\usepackage{todonotes} 
\usepackage{amssymb} 
\usepackage{dsfont} 
\usepackage{braket} 
\usepackage{array} 
\usepackage[unicode,breaklinks]{hyperref}
\hypersetup{
    unicode=true,
    a4paper=true,
    plainpages=false,
    pdftitle={Lattice-depth measurement using continuous grating atom diffraction},
    pdfauthor={B. T. Beswick, S. A. Gardiner, I. G. Hughes},
    pdfsubject={Lattice-depth measurement using continuous grating atom diffraction},
    colorlinks=true,
    linkcolor=blue,
    citecolor=blue,
    filecolor=black,
    urlcolor=blue
}
\hypersetup{colorlinks=true,pdfborder={0 0 0}} 
\usepackage{upgreek} 

\usepackage[caption=false]{subfig} 
\usepackage{graphicx} 
\usepackage[countmax]{subfloat}
\usepackage{import} 
\usepackage{transparent} 
\usepackage[export]{adjustbox}

\definecolor{MyGreen}{rgb}{0,0.5,0} 

\graphicspath{ {../Figures/} }

\begin{document}

\title{Lattice-depth measurement using continuous grating atom diffraction}
\author{Benjamin T. Beswick}\email{b.t.beswick@durham.ac.uk} 
\author{Ifan G. Hughes}\email{i.g.hughes@durham.ac.uk}
\author{Simon A. Gardiner}\email{s.a.gardiner@durham.ac.uk}
\affiliation{Joint Quantum Centre (JQC) Durham--Newcastle,
Department of Physics, Durham University, Durham DH1 3LE, United Kingdom}

\date{\today}

\begin{abstract}
We propose a new approach to characterizing the depths of optical lattices, in which an atomic gas is given a finite initial momentum, which leads to high amplitude oscillations in the zeroth diffraction order which are  robust to finite-temperature effects. We present a simplified model yielding an analytic formula describing such oscillations for a gas assumed to be at zero temperature. This model is extended to include atoms with initial momenta detuned from our chosen initial value, before analyzing the full finite-temperature response of the system. Finally we present a steady-state solution to the finite-temperature system, which in principle makes possible the measurement of both the lattice depth, and initial temperature of the atomic gas simultaneously.
\end{abstract}

\maketitle

\section{Introduction}
There is much interest in the precise measurement of optical lattice \cite{bec_in_optical_lattice_morsch_2006} depths in the field of atomic physics, particularly for accurate determination of transition matrix elements \cite{mitroy_safronova_matrix_elements_2010,arora_tuneout_wavelengths_2011,henson_tuneout_helium_2015,safronova_rb_matrix_elements_2015,clark_bec_matrix_elements_2015},  better knowledge of these matrix elements can be used to improve the black body radiation correction for ultraprecise atomic clocks \cite{safronova_bbr_2011,sherman_polarizability_lattice_clock_2012}, and allows quantitative modeling of atom-light interaction \cite{whiting_et_al_matrix_elements_2016}. Other areas of interest include atom interferometry \cite{Cronin_2009_atom_interferometry,atom_interferometry_2006} and many body quantum physics \cite{bloch_many_body_physics_lattices_2008,gyuboong_kagome_lattice_2012}, where knowledge of the lattice depth is essential for interpreting experimental results.

Commonly used lattice depth measurement schemes include Kapitza--Dirac scattering \cite{cahn_kaptizadirac_1997,gadway_kapitzadirac_2009,birkl_bragg_scattering_optical_lattice_1995,cheiney_matterwave_scattering_2013,gyuboong_kagome_lattice_2012}, parametric heating \cite{friebel_parametric_heating_1998}, Rabi oscillations \cite{ovchinnikov_rabi_oscillations_1999}, and, more recently, the sudden phase shift method \cite{cabreragutierez_sudden_phase_shift_2018}. For the case of a weak lattice ($V\protect{\lesssim}0.01E_\mathrm{R}$ for any atom, where $V$ is the lattice depth and $E_\mathrm{R}$ is the atomic recoil energy), methods based on multipulse atom diffraction have been explored \cite{herold_matrix_elements,kao_tuneout_dysprosium}, with a view to reducing signal-to-noise considerations in the measurement of the resultant diffraction patterns. In previous work we have presented improved models for the expected multipulse diffraction patterns for a given lattice depth. We have also noted that  when considering a gas with initial momentum $\hbar K/2$, the functional form of these models is markedly simpler and therefore easier to fit to data to make an accurate measurement of the lattice depth \cite{beswick_et_al_multipulse}.

In this paper we explore such a measurement scheme for a lattice which is not pulsed but instead continuously present throughout the experimental sequence, which we show to be more robust to finite-temperature effects than a multipulse approach. In Sec.\ \ref{system}, we describe our model system and experimental considerations. In Sec.\ \ref{twostateanalytics}, we introduce a simplified analytic approach for determining the time evolution of the atomic population in the zeroth diffraction order, and make a comparison to exact numerical calculations. Finally, in Sec.\ \ref{finitetemperature}, we present an approximate analytic model for the finite-temperature response of the system, and discuss how these may be used to determine both the lattice depth and initial temperature of the atomic gas.
\section{Model system: Atomic gas in an optical grating}
\label{system}
\subsection{Experimental setup and Hamiltonian}\label{experimentalsetup}
\begin{figure}[!htbp]
\includegraphics{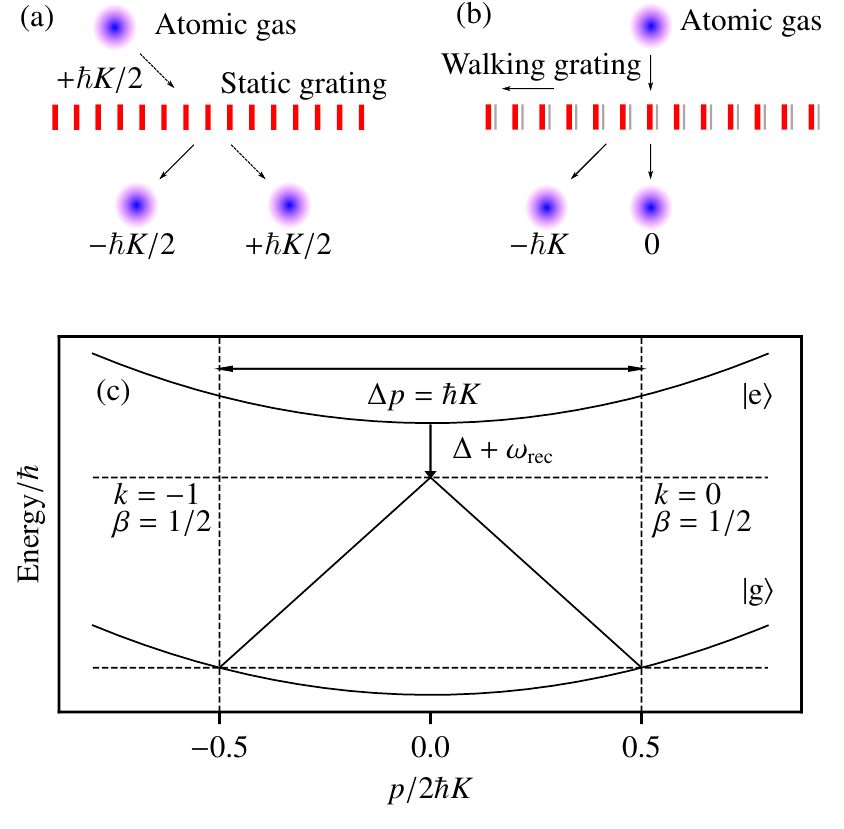}
\caption{(Color online) (a) A BEC initially prepared in the $p=+\hbar K/2$ state, where $K$ is twice the laser wavenumber $k_\mathrm{L}$, is exposed it to a static optical grating, causing it to diffract into an, in principle, infinite number of momentum states separated by integer multiples of $\hbar K$, here we show only the first diffraction order. Equivalently, the BEC may be prepared in the $p=0$ state, and exposed to a walking grating with an linearly time-dependent phase [see Eq.\ (\ref{HlattUntrans})] as in (b). The dynamics of the setup is identical, though the momenta in the lab frame are shifted by $-\hbar K/2$. (c), semiclassical energy-momentum diagram for a single two-level atom scattering photons from a static optical grating. The atom begins on the ground state energy parabola, with classical momentum $p=\hbar K/2$ before scattering a photon carrying momentum $p=-\hbar K/2$ and energy $\hbar^2 K^2/2M$, to reach the detuned virtual state above, before undergoing stimulated emission back to the ground state, resulting in a total momentum transfer of $\Delta p = -\hbar K$. This scattering process and its exact reversal are the only processes which semiclassically conserve both the energy and momentum of the atom grating system, indicating that population transfer between the $p=\hbar K/2$ and $p=-\hbar K/2$ states and vice versa ought to be the dominant process in the system.
\label{fig:systemDiagram}
}
\end{figure}
We consider a two-level atom in an assumed noninteracting Bose-Einstein condensate exposed to a far off resonance optical grating, the Hamiltonian of which is given by Eq.\ (\ref{HlattUntrans}):
\begin{equation}
\tilde{H}_{\mathrm{Latt}} = \frac{\hat{p}^2}{2M} - V \cos \left( K\left[\hat{x}+v_\phi t \right] \right), 
\label{HlattUntrans}
\end{equation}
where $\hat{p}$ is the momentum operator along the lattice axis, $V$ is the lattice depth, $K$ is twice the laser wavenumber $k_\mathrm{L}$, $M$ is the atomic mass and $v_\phi$ is the phase velocity of the grating in the $x$ direction ($v_\phi$=0 for a static grating). For the simpler case of a static grating, we consider a BEC initially prepared in a momentum state with $p=\hbar K/2$.\footnote{The initial momentum $p=\hbar K/2$ is chosen with a view to creating population oscillations between the zeroth and first diffraction orders with a strong sinusoidal character, as suggested in  \cite{beswick_et_al_multipulse}.} As shown in Fig.\ \ref{fig:systemDiagram}(a), the BEC is diffracted by the static optical grating for a time $t$, before a time of flight measurement interrogates the population of the gas in each of the allowed momentum states. In principle there is an infinite ladder of such states, each separated by integer multiples of $\hbar K$ \cite{kicked_rotor_wigner,beswick_et_al}, though here we show only the zeroth and first diffraction orders. We note that an initial state $p=\hbar K/2$ can be achieved for instance by Bragg diffraction, or equivalently we may prepare the BEC in a state with $p=0$ and impart an appropriately tuned time-dependent phase $v_\phi t$ to the standing wave as in Fig.\ \ref{fig:systemDiagram}(b). We show this equivalency in Sec.\ \ref{gaugetransformations} below.
\subsection{Gauge transformations and momentum kicks}\label{gaugetransformations}
The Hamiltonian of Eq.\ (\ref{HlattUntrans}) can be transformed to a frame comoving with the walking grating by use of the unitary transformation
\begin{equation}
\hat{U}=\hat{U}_x\hat{U}_p\hat{U}_\alpha=\exp\left(i m v_\phi \hat{x}/\hbar\right)\exp\left(-i v_\phi \hat{p}t/\hbar\right)\exp\left(i \alpha t/\hbar\right),
\label{Utrans}
\end{equation}
where we have chosen $\alpha=Mv_\phi^2 /2$ for convenience. Using $\hat{U}_p \hat{x}\hat{U}^\dagger_p=\hat{x}+v_\phi t$ and $\hat{U}_x \hat{p}\hat{U}^\dagger_x=\hat{p}+Mv_\phi$. This transformation yields:
\begin{equation}
\hat{H}_{\mathrm{Latt}} = \frac{\hat{p}^2}{2M} - V \cos \left( K\hat{x} \right).
\label{Hlatt}
\end{equation}
The Hamiltonian of Eq.\ (\ref{Hlatt}) describes the system in a frame moving with velocity $-v_\phi$, therefore, a gas moving with velocity $v=0$ in the moving frame appears to move with velocity $-v_\phi$ in the lab frame. Conversely, a gas moving with velocity $v=\hbar K/2M$ in the comoving frame, moves with velocity $v=(\hbar K/2M) -v_\phi$ in the lab frame. Choosing $v_\phi=0$ yields the case in Fig.\ \ref{fig:systemDiagram}(a), while with $v_\phi=\hbar K/2M$, we have the situation shown in Fig.\ \ref{fig:systemDiagram}(b).

The spatial periodicity of Eq.\ (\ref{Hlatt}) allows us to invoke Bloch theory \cite{ashcroft_mermin}, by rewriting the momentum operator in the following basis:
\begin{subequations}
\begin{align}
(\hbar K)^{-1}\hat{p}  & = \hat{k} + \hat{\beta},
\label{Eq:MomentumParam}
\\
\hat{k}|(\hbar K)^{-1}p = k + \beta\rangle 
& = k|(\hbar K)^{-1}p = k + \beta\rangle,
\label{keigenstates}
\\
\hat{\beta}|(\hbar K)^{-1}p = k + \beta\rangle 
& = \beta|(\hbar K)^{-1}p = k + \beta\rangle.
\label{kbetaeigenstates}
\end{align}
\end{subequations}
We may speak of $k \in \mathbb{Z}$ as the discrete part of the momentum, and $\beta \in [-1/2, 1/2)$ as the continuous part or \textit{quasimomentum} \cite{bach_burnett_d'arcy_gardiner_2005}. Here $\beta$ is a conserved quantity, as such, only momentum states separated by integer multiples of $\hbar K$ are coupled \cite{kicked_rotor_wigner,beswick_et_al}. This simplification allows us to construct the time evolution operator for a lattice pulse of duration $t$ from the lattice Hamiltonian (\ref{Hlatt}) as follows:
\begin{equation}
\hat{U}(\beta,\tau)_{\mathrm{Latt}} =
\exp
\left(
-i\left[
\frac{\hat{k}^{2}+2\hat{k}\beta}{2}-V_{\mathrm{eff}}\cos(\hat{\theta})\right]\tau
\right)\,,
\label{floquetfinitedurationk}
\end{equation}
in which $\beta$ is simply a scalar value such that overall phases which depend solely on $\beta$ can be neglected. Here $V_{\mathrm{eff}}= VM/\hbar^2 K^2$ is the dimensionless lattice depth, $\hat{\theta}=K \hat{x}$ and $\tau=t\hbar K^2/ M$ is the rescaled time.

By using Eq.\ (\ref{floquetfinitedurationk}) to calculate $|\psi(\tau) \rangle=\sum_j c_j(\tau) | k = j \rangle$, the population in each discrete momentum state $| k = j \rangle$ following an evolution for a rescaled time of $\tau$ is given by the absolute square of the coefficients $P_j(\tau)=|c_j(\tau)|^2$. In this paper we employ the well-known split-step Fourier approach \cite{daszuta_andersen_2012,beswick_et_al} to determine $|\psi(\tau) \rangle$, as well as an analytic approach based on a simpler two-state model.

The dynamics of a single atom in the BEC standing-wave system can be understood in terms of the scattering process given by the semiclassical energy diagram of Fig.\ \ref{fig:systemDiagram}(c) (see also \cite{martin_bragg_scattering,giltner_bragg_atom_interferometer,borde_in_atom_interferometry_1997,kozuma_splitting_with_bragg_diffraction_1999,gupta_coherent_manipulation_standing_light_wave_2001}). A two-level atom begins in a state with momentum $p=\hbar K/2$, before absorbing a photon with momentum $p=-\hbar K/2$, and subsequently emits a second photon with the momentum $p=\hbar K/2$. This is the only scattering process which classically conserves energy, whilst also conserving the quasimomentum. We therefore expect that scattering into states with momentum $p>|\hbar K/2|$ ought to be strongly suppressed even under the fully quantum time evolution. We explore this simplified picture in Sec.\ \ref{twostateanalytics}.
\section{Reduction to an effective 2-state system}
\label{twostateanalytics}
\subsection{Simplification}
\begin{figure*}[!htbp]
{\centering
\includegraphics{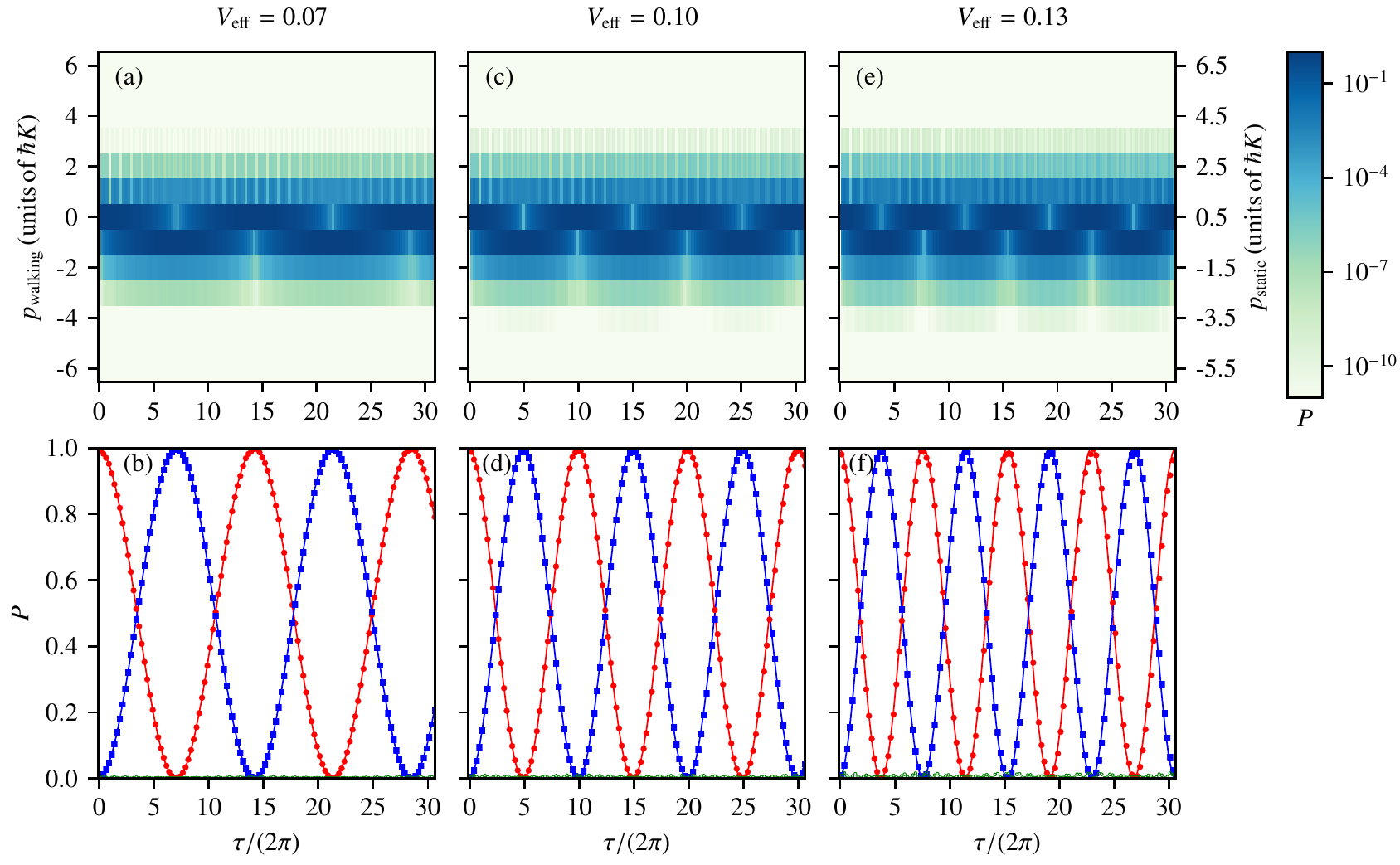}
\caption{(Color online) Time evolved momentum distributions for an atomic gas initially prepared in the $|k=0,\beta=1/2 \rangle$ momentum state (corresponding to the $|k=0,\beta=0 \rangle$ state in the lab frame for a walking grating), as calculated numerically on a basis of 2048 momentum states. The top row of false color plots [(a),(c),(e)] shows the population in the first 13 momentum states, to be read on the logarithmic colorbar to the right, a cutoff population of $P_\mathrm{cutoff}=10^{-11}$ has been applied to accommodate the log scale. The labels $p_\mathrm{static}$ and $p_\mathrm{walking}$ denote the momentum as measured in the lab frame for the case of a static and a walking grating respectively. The bottom row of plots [(b),(d),(f)] shows the time evolution of the population in the $|k=0\rangle$ (red circles) and $|k=-1\rangle$ (blue squares) states, where the solid line through each curve is given by the analytic solution of Eqs.\ (\ref{pzero}) and (\ref{pminusone}). Also shown is the population in the $|k=1\rangle$ state (green points). Each column of plots corresponds to a simulation for a fixed value of the effective lattice depth $V_\mathrm{eff}$, here, from left to right $V_\mathrm{eff}=0.07,0.10,0.13$ respectively.
\label{fig:momentumdistributions} 
}
}
\end{figure*}

We may test the conjecture that population transfer into states with $k<-1$ or $k>0$ is strongly suppressed by computing the full time evolution of the system numerically, the results of such calculations on an exhaustive basis of momentum states are displayed in Fig\ \ref{fig:momentumdistributions}. Over the 13 basis states displayed, we can clearly see that, though population transfer into higher order modes does occur, the oscillation of population between the $k=-1$ and $k=0$ states is the dominant process in the system. We therefore expect that a representation of the system in a truncated momentum basis composed of only these two states ought to capture the essential dynamics, and explore this simplified two-state model below.
\subsection{Two-state model analytics}
\begin{figure}[!htbp]
{\centering
\includegraphics{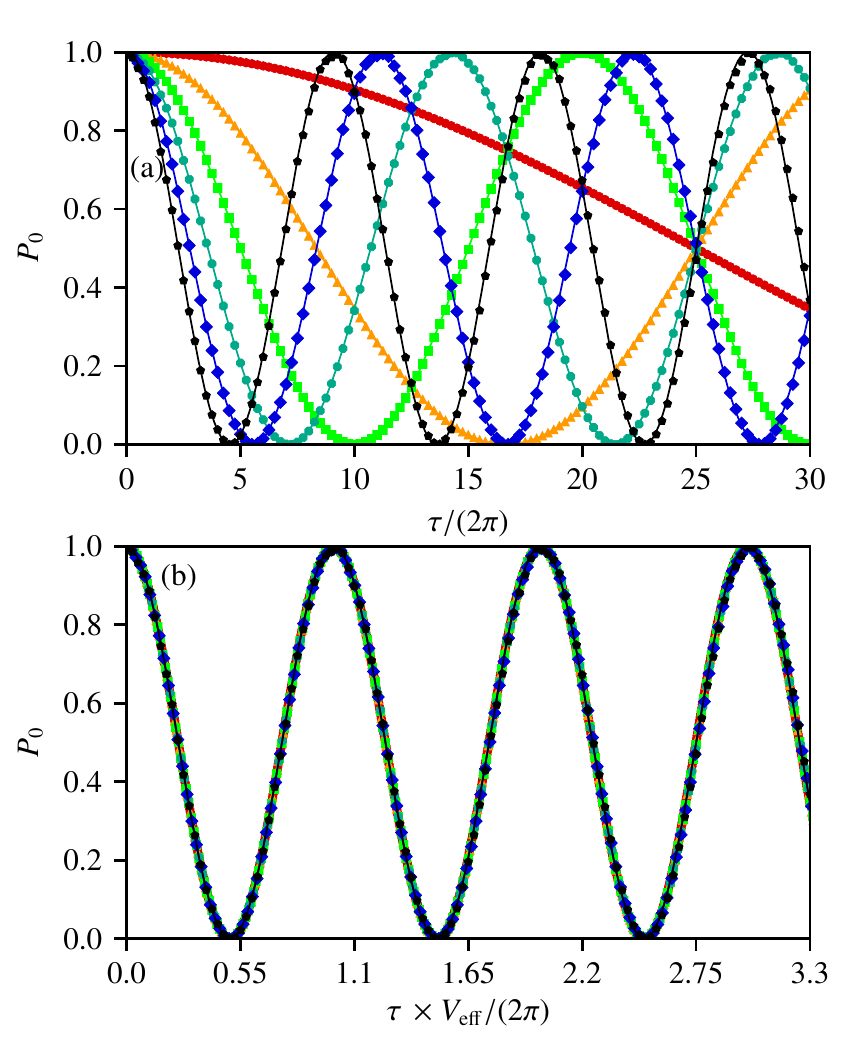}
\caption{(Color online) (a): Plot of $P_0$, the population in the $|k=0\rangle$ state, versus number of pulses, as calculated on a basis of 2048 momentum states using a split-step Fourier method (solid markers). The solid lines correspond to the analytic solution for $P_0$ in a two state basis, as given by Eq.\ (\ref{pzero}). Each set of markers corresponds to a fixed value of the effective lattice depth ranging from the slowest-oscillating curve at $V_\mathrm{eff}=0.01$ to the fastest oscillating one at $V_\mathrm{eff}=0.11$ in steps of $0.02$. (b): Reproduction of (a), with the number of pulses axis scaled by $V_\mathrm{eff}$ to reveal a universal curve both in the analytics and the numerical simulations. The data have been extended to span the full range of the horizontal axis.\label{fig:universalcurve}
}
}
\end{figure}

We may represent the Hamiltonian (\ref{Hlatt}) in the $\beta=1/2$ subspace using the following two-state momentum basis:
\begin{subequations}
\label{trunc_basis}
\begin{align}
|k=0\rangle &= \left( \begin{array}{c}
1\\
0\\
\end{array}
\right), \label{column0} \\
|k=-1\rangle&= \left( \begin{array}{c}
0\\
1
\end{array}
\right), \label{columnminus}
\end{align}
\end{subequations}
yielding:
\begin{equation}
H^{2\times2}_{\mathrm{Latt}}=
\begin{pmatrix}
    1/4  &  -V_\mathrm{eff}/2 \\
      -V_\mathrm{eff}/2  &  1/4 
\end{pmatrix}.
\label{betahalftwostate}
\end{equation}
We recognize Eq.\ (\ref{betahalftwostate}) as a Rabi matrix with zero detuning, the eigenvectors and eigenvalues of which are well known \cite{barnett_radmore_methods_qo_1997}, and can be used to straightforwardly determine the time evolution of the population in the $|k=0\rangle$ and $|k=-1\rangle$ states, respectively:
\begin{subequations}
\begin{align}
\label{pzero}
P&_0=\cos^2(V_\mathrm{eff}\tau/2), \\
\label{pminusone}
P&_{-1}=\sin^2(V_\mathrm{eff}\tau/2),
\end{align}
\end{subequations}
as outlined in Appendix \ref{appendixtwostate}. This analytic result is compared to our exact numerics in Figs.\ \ref{fig:momentumdistributions} and \ref{fig:universalcurve}, both of which show excellent agreement for a wide range of experimentally relevant values of the effective lattice depth $V_\mathrm{eff}$. We note in particular that the form of Eqs.\ (\ref{pzero}) and (\ref{pminusone}) is such that there is an exact universality between $\tau$ and $V_\mathrm{eff}$, which is elucidated in Fig.\ \ref{fig:universalcurve}(b), where all population curves fall on top of each other.
\section{Finite-temperature response}
\label{finitetemperature}
\subsection{Other values of $\mathbf{\beta}$}
\begin{figure}[!htbp]
{\centering
\includegraphics{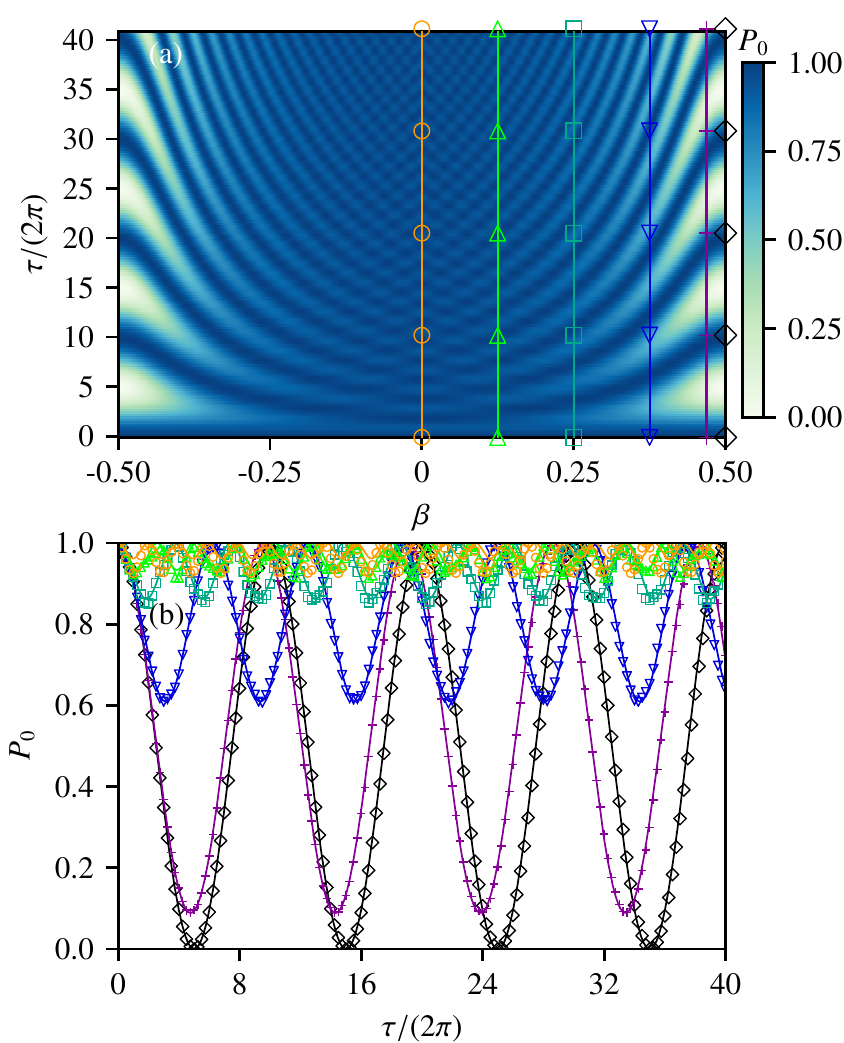}
\caption{(Color online) (a): False-color plot of the time evolution of $P_0$ as computed in a basis of 2048 momentum states for values of the dimensionless quasimomentum $\beta$ [see Eqs.\ (\ref{keigenstates},\ref{kbetaeigenstates})] ranging from $\beta = -0.5$ to $\beta = 0.5$ in steps of $\beta =0.00025$ (4001 quasimomentum values). We have chosen a relatively large lattice depth of $V_\mathrm{eff}=0.1$ such that the different dynamical behaviors are made clear for the chosen evolution time $\tau/(2\pi)=40$\label{betacarpet}. (b): Slices taken through the quasimomentum distribution parallel to the time axis for $\beta=0,0.0625,0.125$, then increasing in increments of $\beta=0.125$ up to a maximum of $\beta=0.5$, enclosing the full range of dynamics in the $k=0$ subspace. Each vertical set of markers in (a) corresponds to the position in the quasimomentum distribution of the slices in (b), where the solid lines represent our analytic solution for each $\beta$ subspace [Eq.\ (\ref{pzerobetatau})].\label{betaplanewaves}
}
}
\end{figure}

In the following section we consider the effect of evolving initial states with quasimomentum different to $\beta=1/2$ in order to gain insight into the dynamics of a finite-temperature gas. Numerically, this is achieved by computing the evolution of an initial state $|k+\beta\rangle$ under the time evolution operator (\ref{floquetfinitedurationk}). We make the assumption from the outset that the initial momentum distribution of the gas (centred at $\beta=1/2$) spans less than half of each of the $k=0$ and $k=-1$ Brillouin zones for a static grating (or falls within the $k=0$ Brillouin zone with a momentum distribution centered on $\beta=0$ for a walking grating). Our results in this low temperature regime are displayed in Fig.\ \ref{betaplanewaves}, which indicates a $k=0$ Brillouin zone with high amplitude but low-frequency oscillations in the population of the zeroth diffraction order centered around $|\beta|=1/2$, and low amplitude but rapidly oscillating solutions as $\beta$ is detuned from this value.
We may also use our simplified semiclassical model of Sec.\ \ref{twostateanalytics} to derive an approximate analytic result for the same calculation, in which the quasimomentum $\beta$ is encoded as a detuning to be included in our initial Rabi model of Eq.\ (\ref{betahalftwostate}). These additions yield the following $2\times2$ Hamiltonian matrix:
\begin{equation}
H^{2\times2}_{\mathrm{Latt}}(\beta)=
\begin{pmatrix}
    \beta^2/2  &  -V_\mathrm{eff}/2 \\
   -V_\mathrm{eff}/2  &  (1-2\beta + \beta^2)/2 
\end{pmatrix},
\label{betatwostate}
\end{equation}
in which $\beta$ is now a free parameter. The time evolution of the zeroth diffraction order population governed by this matrix can be found using the approach given in Appendix \ref{appendixbeta}, thus:
\begin{equation}
P_0(\beta)=
1-\frac{V_\mathrm{eff}^2}{(\beta-1/2)^2 + V_\mathrm{eff}^2}\sin^2\left(\sqrt{(\beta -1/2)^2 + V_\mathrm{eff}^2}\frac{\tau}{2}\right),
\label{pzerobetatau}
\end{equation}
which is similar to the result reported in \cite{gadway_kapitzadirac_2009} for a zero temperature gas, and agrees excellently with the exact numerics for physically relevant parameters as shown in Fig.\ \ref{betaplanewaves}. We therefore expect that thermal averaging of this result should produce an accurate description of the full finite-temperature response.
\subsection{Finite temperature analysis}
\begin{figure*}[!htbp]
{\centering
\includegraphics{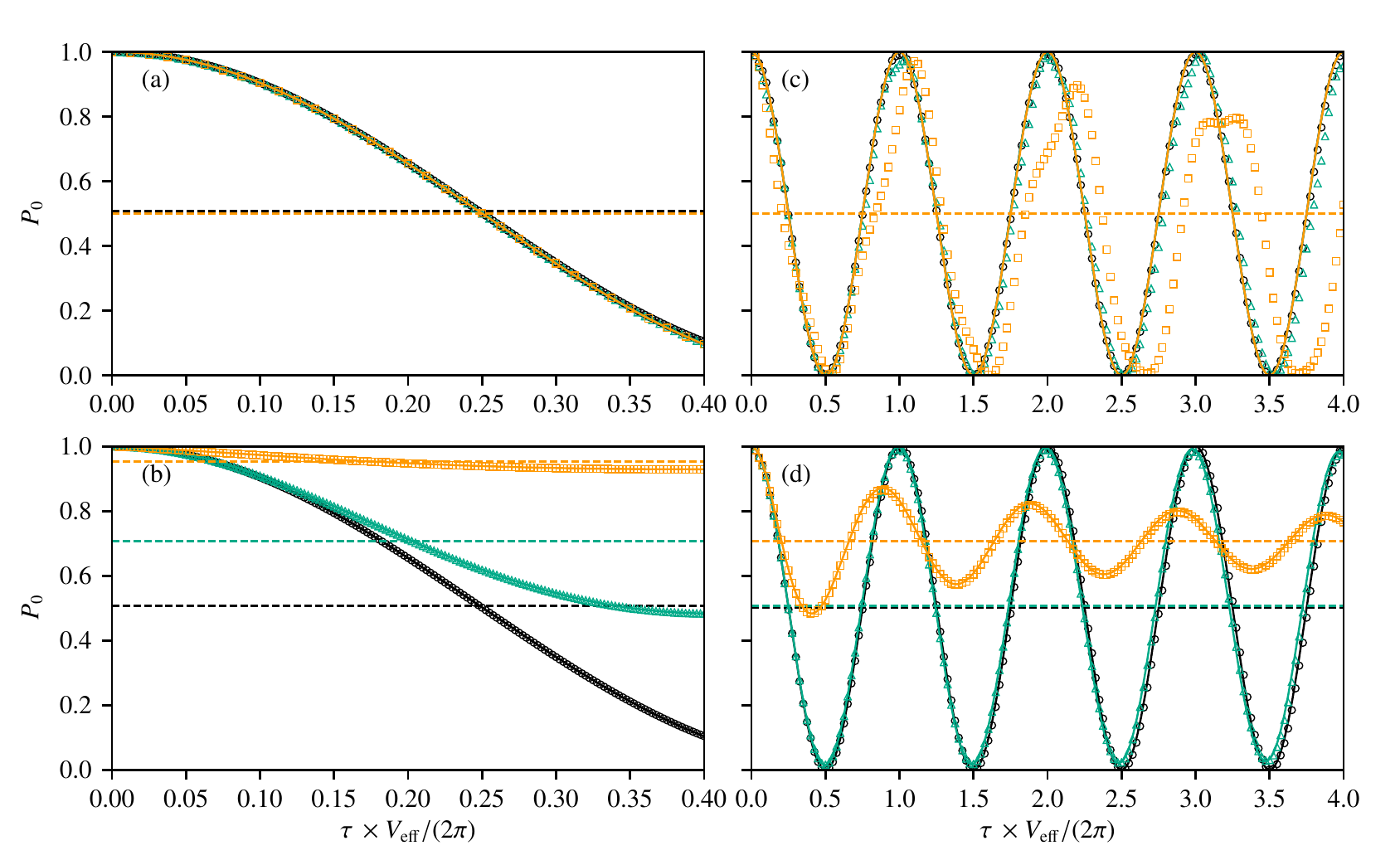}
\caption{(Color online) Plot of the finite temperature response of $P_0$ vs (number of pulses)$\times V_\mathrm{eff}$, where $V_\mathrm{eff}$ is the dimensionless lattice depth [see Eq.\ (\ref{floquetfinitedurationk})], as calculated for an ensemble of 4001 particles each evolved in a basis of 2048 momentum states (hollow markers). The left column [(a), (b)] corresponds to the weak-lattice regime, and the right column [(c), (d)] to the strong-lattice regime. The top row of plots [(a), (c)] shows the finite-temperature response of $P_0$ at a temperature of $w=0.00125$ for a selection of different lattice depths, $V_\mathrm{eff}=0.01,0.02,0.05$ (all curves fall on top of each other) in the weak regime (a) and $V_\mathrm{eff}=0.1,0.2,0.5$ (lower, middle and uppermost curves) in the strong regime (b). For the bottom row [(b), (d)], each set of curves and markers corresponds to the response of $P_0$ at a different temperature ($w=0.00125,0.0125,0.125$; lower, middle and uppermost curves respectively), where the effective lattice depth is kept constant at $V_\mathrm{eff}=0.1$ in the strong-lattice case and $V_\mathrm{eff}=0.01$ the weak-lattice case. In all panels, the solid lines represent the result yielded by numerically integrating Eq.\ (\ref{integralequationrescaled}). The horizontal dashed lines correspond to the result of the steady state solution of Eq.\ (\ref{steadystate}) for each set of parameters.\label{finitetempresponse}
}
}
\end{figure*}
To find the finite temperature response of the system we weight the contribution of Eq.\ (\ref{pzerobetatau}) for each individual quasimomentum subspace according to the Maxwell-Boltzmann distribution:
\begin{equation}
D_{k=0}(\beta,w)=\frac{1}{w \sqrt{2\pi}} \exp \left(\frac{-(\beta-1/2)^2}{2 w^2} \right),
\label{gaussiandist}
\end{equation}
where the dimensionful temperature is given by $\mathcal{T}_{w}=\hbar^{2} K^{2} w^{2}/M k_{\mathrm{B}}$ \cite{saunders_halkyard_challis_gardiner_2007}. Mathematically this corresponds to the integral:
\begin{equation}
P_0(w)=\int_0^1 D_{k=0}(\beta,w)P_0(\beta)\,\mathrm{d}\beta.
\label{integralequation}
\end{equation}
Inserting Eqs.\ (\ref{gaussiandist}) and (\ref{pzerobetatau}), we have:
\begin{equation}
\begin{split}
P_0&(\rho)=\\
&\frac{1}{\sqrt{2\pi}\rho}\int_\frac{-1}{2}^\frac{1}{2}\exp\left(\frac{-\gamma^2}{2\rho^2}\right)\left[ 1-\frac{1}{\gamma^2 + 1}\sin^2\left(\frac{\sqrt{\gamma^2 + 1}}{2}\phi\right)\right]\mathrm{d}\gamma,
\label{integralequationrescaled}
\end{split}
\end{equation}
where we have introduced $\gamma=(\beta-1/2)/V_\mathrm{eff}$, $\phi=V_\mathrm{eff}\tau$ and $\rho=w/V_\mathrm{eff}$ for simplicity. 
The exponential and trigonometric terms can be power expanded, and the integral (\ref{integralequationrescaled}) solved term by term, giving:
\begin{equation}
P_0(\rho)=1-\sum_{s=0}^\infty \sum_{q=0}^s u_s(\phi) M_{s,q} v_q(\rho),
\label{matrixequation}
\end{equation}
where $u_s(\phi)=(-\phi^2)^{s+1}s!/[2(s+1)]!$, $M_{s,q}=\protect{-(2q)!/[2(q!)^2(s-q)!]}$ and $v_q(\rho)=(\rho^2/2)^q$ (see Appendix \ref{matrixappendix}). Equation (\ref{matrixequation}) can in principle be solved numerically by recursively populating the elements of a sufficiently large pair of $u(\phi)$, $v(\rho)$ vectors and $M$ matrix, though the elements of the vectors will grow with $s$ and $q$ respectively unless $\phi$ and $\rho$ are sufficiently small, and this condition is only satisfied for certain experimentally relevant regimes. Nonetheless, Eq.\ (\ref{matrixequation}) yields some insight when expressed as a sum over derivatives of sinc functions (see Appendix \ref{sincfunctionsappendix}):
\begin{equation}
\begin{split}
P_0&(\rho)=\\
&1-\sum_{q=0}^\infty \left(\frac{\rho}{2}\right)^{2q}\frac{(2q)!}{q!^2}\left\{\left(\frac{\phi}{2}\right)^{2(q+1)} \left[ \left(\frac{2}{\phi}\right)\frac{\mathrm{d}}{\mathrm{d}(\phi/2)}\right]^q \left[\frac{\sin^2(\phi/2)}{(\phi/2)^2} \right] \right\}.
\label{sincderivs}
\end{split}
\end{equation}
With $q=0$, Eq. (\ref{sincderivs}) reduces to the zero temperature result of Eq.\ (\ref{pzero}), as such we should expect the finite temperature behavior of the system to be captured in terms with $q>0$. Though the full sum over $q$ is always convergent, the presence of the $(\phi/2)^{2(q+1)}$ term guarantees that all individual terms with $q\geq 1$ diverge, meaning that a preferred truncation of the sum is not obvious.

However, given the well-behaved nature of the integrand, Eq.\ (\ref{integralequationrescaled}) can be straightforwardly integrated numerically, for instance using the trapezium rule. We compare this numerical integration to our full finite-temperature numerics in Fig.\ \ref{finitetempresponse}, which shows excellent agreement across a large range of initial momentum widths in the weak lattice regime [Figs.\ \ref{finitetempresponse} (a),(b)], and for $V_\mathrm{eff}=0.1$ in the strong lattice regime [Fig.\ \ref{finitetempresponse} (d)]. However, for $V_\mathrm{eff}=0.5$ [Fig.\ \ref{finitetempresponse} (c)] the agreement is relatively poor, as in this regime the semiclassically motivated two-state model is no longer valid. We therefore expect that numerically fitting Eq.\ (\ref{integralequationrescaled}) to experimental data, with $\phi=V_\mathrm{eff}\tau$ and $\rho=w/V_\mathrm{eff}$ as free parameters, would give an accurate value of the effective lattice depth, if the time $\tau$ is known to high precision and the lattice depth is sufficiently small.

Further, we note that using standard integral results, we may also extract the steady state solution to Eq.\ (\ref{integralequationrescaled}) as $\protect{\phi\rightarrow\infty}$:
\begin{equation}
P_{0,\phi\rightarrow\infty}(\rho)=\frac{1}{2\rho}\sqrt{\frac{\pi}{2}}\exp\left(\frac{1}{2\rho^2}\right)\mathrm{Erfc}\left(\frac{1}{\sqrt{2}\rho}\right),
\label{steadystate}
\end{equation}
which depends only on $\rho=w/V_\mathrm{eff}$. Here, `Erfc' is the complementary error function \cite{hughes_hase_measurements_uncertanties_2010}.\footnote{When evaluating Eq.\ (\ref{steadystate}) for physically relevant values of $\rho=w/V_\mathrm{eff}$, the exponential term becomes large as the error function takes a correspondingly such that $P_{0,\phi\rightarrow\infty}(\rho)$ remains bounded between 0 and 1. This complication can present a problem for numerical evaluation using standard numerical routines. In practice, we numerically implement Eq.\ (\ref{steadystate}) exclusively in terms of rational numbers in Mathematica, before requesting a numerical evaluation to a specified precision.} In essence, by measuring the steady state population experimentally, and numerically fitting Eq.\ (\ref{steadystate}), $\rho=w/V_\mathrm{eff}$ can be straightforwardly determined and substituted into Eq.\ (\ref{integralequationrescaled}), leaving a fit in only one parameter $\phi=V_\mathrm{eff}\tau$. The steady state population can be found either by allowing the atomic gas to evolve in the lattice for a sufficiently long time, or taking the average value of $P_0$ in time for an appropriate number of oscillations. In fact, this improved fitting approach not only allows $\phi=V_\mathrm{eff}\tau$, and therefore the effective lattice depth $V_\mathrm{eff}$ to be determined more accurately, but also allows the initial effective temperature to be determined from $w=\rho V_\mathrm{eff}$.
\section{Conclusions}
\label{conclusion}
We have presented a simplified model system yielding an analytic zero-temperature formula for the evolution of the zeroth diffraction order population, and demonstrated the validity of this approach across a wide range of lattice depths. We have extended this model to incorporate finite-temperature effects and discussed from where they arrive mathematically. We have shown that there is excellent agreement between this analytic model and exact numerical calculations if the lattice depth is sufficiently small, and shown that a steady state solution exists, which may be useful for determining the lattice depth and initial temperature of a gas from a single set of population measurements. With regard to potential experimental implementations, we note that the phase velocity of a walking optical lattice can be calibrated extremely precisely, however, does require optical elements to be in place which will reduce the intensity of the laser beam and therefore the lattice. The alternative is to impart a specified momentum to an initially stationary BEC; it is unlikely that this can be achieved with the same level of precision, however there is no need for any additional optical elements affecting the lattice depth.
\acknowledgments
B.T.B., I.G.H., and S.A.G. thank the Leverhulme Trust research program grant RP2013-k-009, SPOCK: Scientific
Properties of Complex Knots for support. We would also like to acknowledge helpful discussions with Andrew R. MacKellar.
\appendix
\section{Derivation of the two-state model}\label{appendixtwostate}
To calculate the time-evolution of the population in the zeroth diffraction order, we construct the time evolution operator in the momentum basis from the Hamiltonian of Eq.\ (\ref{betahalftwostate}), reproduced here for convenience:
\begin{equation}
H^{2\times2}_{\mathrm{Latt}}=
\begin{pmatrix}
    1/4  &  -V_\mathrm{eff}/2 \\
   -V_\mathrm{eff}/2  &  1/4 
\end{pmatrix}.
\end{equation}
The diagonal terms simply represent an energy shift that can be transformed away, thus the eigenvalues of Eq.\ (\ref{betahalftwostate}) can simply be read from the off-diagonal: $E_\pm=\pm V_\mathrm{eff}/2$. We may now solve the eigenvalue equation:
\begin{equation}
\begin{pmatrix}
    0  &  -V_\mathrm{eff}/2 \\
   -V_\mathrm{eff}/2  &  0 
\end{pmatrix}
\begin{pmatrix}
v_1^\pm \\
v_0^\pm
\end{pmatrix}
 =\pm V_\mathrm{eff}/2
\begin{pmatrix}
v_1^\pm \\
v_0^\pm
\end{pmatrix}.
\label{eigenvalueeqn}
\end{equation}
Equation (\ref{eigenvalueeqn}) leads directly to $-v_1^\pm=\pm v_0^\pm$, yielding eigenvectors:
\begin{equation}
|E_+\rangle = \frac{1}{\sqrt{2}} \left( \begin{array}{c}
1 \\
-1
\end{array}
\right), 
\qquad
|E_-\rangle = \frac{1}{\sqrt{2}} \left( \begin{array}{c}
1 \\
1
\end{array}
\right). \label{eigenkets}
\end{equation}
We may now construct our initial condition in the energy basis, in which the matrix representation of the time evolution operator
\begin{equation}
\hat{U}(\tau)=\exp\left(-i \hat{H}_{\mathrm{Latt}} \tau \right)
\end{equation}
is diagonal:
\begin{equation}
|\psi(\tau=0)\rangle=|k=0\rangle = \frac{1}{\sqrt{2}}\left(|E_+\rangle + |E_-\rangle \right).
\end{equation}
The time evolution of the population in the zeroth diffraction order is given by:
\begin{align}
P_0&=\left|\frac{1}{2}\left( \langle E_+| + \langle E_- | \right) \hat{U}(\tau) \left( |E_+\rangle + |E_- \rangle \right) \right|^2,
\nonumber
\\
&=\frac{1}{4}\left|e^{-i E_+ \tau} + e^{-i E_- \tau}\right|^2,
\nonumber
\\
&=\frac{1}{4}\left|e^{-i V_\mathrm{eff} \tau/2} + e^{i V_\mathrm{eff} \tau/2}\right|^2,
\nonumber
\\
&=\cos^2(V_\mathrm{eff}\tau/2),
\end{align}
which corresponds to Eq.\ (\ref{pzero}).
\section{Derivation of $\beta$ dependent two-state model}\label{appendixbeta}
To calculate the time-evolved population for a given quasimomentum subspace, we follow the same procedure as in Appendix \ref{appendixtwostate}. Equation (\ref{betatwostate}), reproduced here for convenience
\begin{equation}
H^{2\times2}_{\mathrm{Latt}}(\beta)=
\begin{pmatrix}
    \beta^2/2  &  -V_\mathrm{eff}/2 \\
   -V_\mathrm{eff}/2  &  (1-2\beta + \beta^2)/2 
\end{pmatrix},
\nonumber
\end{equation}
is nothing other than a Rabi matrix, the eigenvalues of which are $E_\pm=\left[(1/2-\beta+\beta^2) \pm \sqrt{(\beta-1/2)^2+V_\mathrm{eff}^2}\right]/2$, and the corresponding eigenvectors:
\begin{subequations}
\begin{align}
\begin{split}
|E_+\rangle 
= 
\left( \begin{array}{c}
\cos(\alpha/2) \\
\sin(\alpha/2)
\end{array}
\right)
&
=
\frac{1}{\sqrt{2}}\left[\sqrt{1+\cos(\alpha)}|k=0\rangle\right. \\
              &\qquad \left. +\sqrt{1-\cos(\alpha)}|k=-1\rangle\right],
\end{split}
\\
\begin{split}
|E_+\rangle 
= 
\left( \begin{array}{c}
-\sin(\alpha/2)
\\
\cos(\alpha/2)
\end{array}
\right)
&
=
-\frac{1}{\sqrt{2}}\left[\sqrt{1-\cos(\alpha)}|k=0\rangle\right. \\
              &\qquad \left. -\sqrt{1+\cos(\alpha)}|k=-1\rangle\right],
\end{split}
\end{align}
\end{subequations}
where $\cos(\alpha)=(\beta-1/2)/\sqrt{(\beta-1/2)^2+V_\mathrm{eff}^2}$. This leads directly to:
\begin{align}
|\psi(\tau=0)\rangle &=|k=0\rangle = \cos(\alpha/2)|E_+\rangle - \sin(\alpha/2)|E_- \rangle
\nonumber
\\
&=\frac{1}{\sqrt{2}}
\left[
\sqrt{1+\cos(\alpha)} |E_+\rangle - \sqrt{1-\cos(\alpha)} |E_-\rangle
\right].
\nonumber
\end{align}
We may now simply calculate the time-evolved state from the action of the time evolution operator
\begin{equation}
\hat{U}(\tau,\beta)=\exp\left(-i \hat{H}(\beta)_{\mathrm{Latt}}\tau\right),
\nonumber
\end{equation}
on this initial state thus:
\begin{align}
|\psi(\tau,\beta)\rangle &= \exp\left(-i \hat{H}(\beta)_{\mathrm{Latt}}\tau \right)|k=0\rangle
\nonumber
\\
&
=
\frac{1}{\sqrt{2}}
\left[
\sqrt{1+c}\, e^{-i E_+ \tau} |E_+\rangle + \sqrt{1-c}\, e^{-i E_- \tau} |E_-\rangle
\right].
\nonumber
\end{align}
Here we have introduced $c \equiv \cos(\alpha)$. The time-evolved population in the zeroth diffraction order for a given $\beta$ subspace is then given by:
\begin{align}
p_0(\tau,\beta)&=|\langle k=0 | \psi(\tau,\beta)\rangle|^2
\nonumber
\\
&=
\frac{1}{4}\left|(1+c)\, e^{-i E_+ \tau} + (1-c)\, e^{-i E_- \tau}\right|^2
\nonumber
\\
&=
\frac{1}{4}\left|
e^{E_+\tau/2}
e^{E_-\tau/2}
\left[
(1+c)\, e^{-i [E_+-E_-] \tau/2} + (1-c)\, e^{i [E_+-E_-] \tau/2}
\right]
\right|^2
\nonumber
\\
&=
\cos^2([E_+-E_-]\tau/2)+c^2\sin^2([E_+-E_-]\tau/2)
\nonumber
\\
&=
1+(c^2-1)\sin^2([E_+-E_-]\tau/2)
\nonumber
\\
&=
1-\frac{V_\mathrm{eff}^2}{(\beta-1/2)^2+V_\mathrm{eff}^2}\sin^2\left(\sqrt{(\beta-1)^2+V_\mathrm{eff}^2}\tau/2\right),
\end{align}
which corresponds to Eq.\ (\ref{pzerobetatau}).
\section{Derivation of finite-temperature matrix equation\label{matrixappendix}}
To derive the matrix equation for the finite-temperature response of the zeroth diffraction order population, we begin from Eq.\ (\ref{integralequation}), into which we insert Eqs.\ (\ref{gaussiandist}) and (\ref{pzerobetatau}), yielding:
\begin{align}
\begin{split}
P_0(w)&=1-\\
&\frac{1}{\sqrt{2\pi}w}\int^\infty_{-\infty} \mathrm{d}\alpha \frac{V_\mathrm{eff}^2}{\alpha^2+V_\mathrm{eff}^2}\exp\left(\frac{-\alpha^2}{2w^2}\right)\sin^2\left(\sqrt{\alpha^2+V_\mathrm{eff}^2}\frac{\tau}{2}\right),
\end{split}
\nonumber
\\
&=1-P_{-1}(w)
\end{align}
where we have introduced $\alpha\equiv(\beta-1/2)$. For simplicity, we now refer to $P_{-1}(w)$, the population in the $|k=-1\rangle$ state. The sinusoidal term can be rewritten using $\sin^2(\theta)=[1-\cos(2\theta)]/2$, thus:
\begin{align}
\begin{split}
P_{-1}(w)=\frac{V_\mathrm{eff}^2}{\sqrt{2\pi}w}\int^\infty_0 \mathrm{d}\alpha\frac{1}{\alpha^2+V_\mathrm{eff}^2}&\exp\left(\frac{-\alpha^2}{2w^2}\right)
\\
&\times\left[1-\cos\left(\sqrt{\alpha^2+V_\mathrm{eff}^2}\tau\right)\right],
\end{split}
\end{align}
where we have used the fact that the integrand is an even function. The term in $\cos\left(\sqrt{\alpha^2+V_\mathrm{eff}^2}\tau\right)$ can then be power expanded, leading to:
\begin{align}
P_{-1}(w)&=\frac{V_\mathrm{eff}^2}{\sqrt{2\pi}w}\int^\infty_0\mathrm{d}\alpha \frac{-1}{\alpha^2+V_\mathrm{eff}^2}\exp\left(\frac{-\alpha^2}{2w^2}\right)\sum_{s=1}^\infty\frac{(-1)^s(\alpha^2+V_\mathrm{eff}^2)^s \tau^{2s}}{(2s)!},
\nonumber
\\
&=\frac{V_\mathrm{eff}^2}{\sqrt{2\pi}w}\sum_{s=0}^\infty\frac{(-1)^s \tau^{2(s+1)}}{(2[s+1])!}\int^\infty_0\mathrm{d}\alpha \exp\left(\frac{-\alpha^2}{2w^2}\right)(\alpha^2+V_\mathrm{eff}^2)^s,
\end{align}
such that the square root in the argument no longer appears, and the $(\alpha^2+V_\mathrm{eff}^2)^s$ term can be binomially expanded thus:
\begin{align}
\begin{split}
P_{-1}(w)=
\\
\frac{V_\mathrm{eff}^2}{\sqrt{2\pi}w}&\sum_{s=0}^\infty\frac{(-1)^s \tau^{2(s+1)}s!}{(2[s+1])!}\sum_{q=0}^s \frac{V^{2(s-q)}}{q!(s-q)!} \int^\infty_0\mathrm{d}\alpha\,\alpha^{2q} \exp\left(\frac{-\alpha^2}{2w^2}\right).
\label{binomialexpansion}
\end{split}
\end{align}
Further, introducing $\xi\equiv\alpha^2/(2w^2)$, the remaining integral can be rewritten as:
\begin{align}
\int^\infty_0\mathrm{d}\alpha\,\alpha^{2q} \exp\left(\frac{-\alpha^2}{2w^2}\right)&=w^{2q+1}2^{q-1/2}\int^\infty_0\mathrm{d}\xi\exp(-\xi)\xi^{q-1/2},
\nonumber
\\
&=w^{2q+1}2^{q-1/2}\Gamma(q+1/2),
\nonumber
\end{align}
which, when substituted into Eq.\ (\ref{binomialexpansion}) leads to:
\begin{align}
\begin{split}
P_{-1}(w)&=\\
\frac{1}{2\sqrt{\pi}}&\sum_{s=0}^\infty\frac{(-1)^s (V_\mathrm{eff} \tau)^{2(s+1)}s!}{(2[s+1])!}\sum_{q=0}^s \frac{1}{q!(s-q)!}\left(\frac{2w^2}{V_\mathrm{eff}^2}\right)^q\Gamma(q+1/2).
\label{binomialexpansiongammafunction}
\end{split}
\end{align}
Finally, noting that $\Gamma(s+1/2)=(2s)!\sqrt{\pi}/(2^{2s}s!)$, Eq.\ (\ref{binomialexpansiongammafunction}) can be rewritten, thus:
\begin{align}
P_{-1}(w)&=\sum_{s=0}^\infty\sum_{q=0}^s\frac{(-V_\mathrm{eff}^2 \tau^2)^{s+1}s!}{(2[s+1])!} \frac{\left(-\frac{1}{2}\right)(2q)!}{(q!)^2(s-q)!}\left(\frac{w^2}{2V_\mathrm{eff}^2}\right)^q,
\label{matrixequationsumpminusone}
\\
&=u_s(V_\mathrm{eff}\tau)M_{s,q}v_q(w/V_\mathrm{eff}),
\nonumber
\end{align}
or, equivalently, with $\phi=V_\mathrm{eff}\tau$ and $\rho=w/V_\mathrm{eff}$:
\begin{equation}
P_0(\rho)=1-P_{-1}(\rho)=1-\sum_{s=0}^\infty \sum_{q=0}^su_s(\phi)M_{s,q}v_q(\rho),
\nonumber
\end{equation}
which corresponds to Eq.\ (\ref{matrixequation}).
\section{Expression of Eq.\ (\ref{matrixequation}) in terms of Sinc functions\label{sincfunctionsappendix}}
Equation (\ref{matrixequationsumpminusone}) can be rewritten as:
\begin{align}
P_{-1}(\rho)&=
\nonumber
\\
&\sum_{q=0}^\infty\left(\frac{\rho^2}{2}\right)^q\frac{(2q)!}{q!^2}\left\{\left(\frac{-1}{2}\right)\sum_{s=q}^\infty\frac{s!}{(2[s+1])!(s-q)!}(-\phi^2)^{s+1}\right\},
\nonumber
\end{align}
where we have used $\phi=V_\mathrm{eff}t$ and $\rho=w/V_\mathrm{eff}$. We now introduce $\tau=\phi^2$ and re-index the sum in $s$, yielding:
\begin{align}
P_{-1}(\rho)&=
\nonumber
\\
&\sum_{q=0}^\infty\left(\frac{\rho^2}{2}\right)^q\frac{(2q)!}{q!^2}\left\{\left(\frac{-1}{2}\right)\sum_{s=q+1}^\infty\frac{(s-1)!}{(2s)!(s-1-q)!}(-1)^s\tau^s\right\}.
\nonumber
\end{align}
Expanding the factorial terms in $s$ and rearranging in $\tau$ in the following way:
\begin{align}
P_{-1}(\rho)&=
\nonumber
\\
\sum_{q=0}^\infty&\left(\frac{\rho^2}{2}\right)^q\frac{(2q)!}{q!^2}\left\{\tau^{q+1}\left(\frac{-1}{2}\right)\sum_{s=1}^\infty\frac{(s-1)(s-2)...(s-q)}{(2s)!}(-1)^s\tau^{s-q-1}\right\},
\nonumber
\end{align}
which we recognize can be expressed as a derivative in $q$, thus:
\begin{equation}
P_{-1}(\rho)=\sum_{q=0}^\infty\left(\frac{\rho^2}{2}\right)^q\frac{(2q)!}{q!^2}\left\{\tau^{q+1}\left(\frac{-1}{2}\right)\frac{\mathrm{d}^q}{\mathrm{d}\tau^q}\sum_{s=1}^\infty\frac{(-1)^s\tau^{s-1}}{(2s)!}\right\}.
\label{sumderivative}
\end{equation}
Equation (\ref{sumderivative}) can be rewritten:
\begin{equation}
P_{-1}(\rho)=\sum_{q=0}^\infty\left(\frac{\rho^2}{2}\right)^q\frac{(2q)!}{q!^2}\left\{\tau^{q+1}\frac{\mathrm{d}^q}{\mathrm{d}\tau^q}\left(\frac{1}{\tau}\left[-\frac{1}{2}\sum_{s=1}^\infty\frac{(-1)^s\tau^{s-1}}{(2s)!}\right]\right)\right\},
\nonumber
\end{equation}
such that the sum in $s$ can now be recognized as a sinusoidal term, yielding:
\begin{equation}
P_{-1}(\rho)=\sum_{q=0}^\infty\left(\frac{\rho^2}{2}\right)^q\frac{(2q)!}{q!^2}\left\{\tau^{q+1}\frac{\mathrm{d}^q}{\mathrm{d}\tau^q}\left(\frac{1}{\tau}\left[\frac{\sin^2(\sqrt{\tau}/2)}{\tau}\right]\right)\right\}.
\nonumber
\end{equation}
Reintroducing $\phi$ leads to: 
\begin{align}
P_{-1}(\rho)&=\sum_{q=0}^\infty\left(\frac{\rho^2}{2}\right)^q\frac{(2q)!}{q!^2}\left\{\phi^{2(q+1)}\left(\frac{1}{2\phi}\frac{\mathrm{d}}{\mathrm{d}\phi}\right)^q\left[\frac{\sin^2(\phi/2)}{\phi^2}\right]\right\},
\nonumber
\\
&=
\frac{1}{2}\sum_{q=0}^\infty\left(\frac{\rho^2}{2}\right)^q\frac{(2q)!}{q!^2}\left\{\left(\frac{\phi^2}{2}\right)^{q+1}\left(\frac{1}{\phi}\frac{\mathrm{d}}{\mathrm{d}\phi}\right)^q\left[\frac{\sin^2(\phi/2)}{\phi^2}\right]\right\},
\nonumber
\\
\begin{split}
&=
\frac{1}{2}\sum_{q=0}^\infty\left(\frac{\rho^2}{2}\right)^q\frac{(2q)!}{q!^2}\\
&\times\left\{\left(\frac{\phi^2}{2}\right)^{q+1}\frac{1}{2^{2q}}\left[\left(\frac{2}{\phi}\right)\frac{\mathrm{d}}{\mathrm{d}(\phi/2)}\right]^q\left[\frac{\sin^2(\phi/2)}{\phi^2}\right]\right\}.
\end{split}
\nonumber
\end{align}
Equivalently,
\begin{equation}
\begin{split}
P_0(\rho)&=1-P_1(\rho)\\
&=1-\sum_{q=0}^\infty \left(\frac{\rho}{2}\right)^{2q}\frac{(2q)!}{q!^2}\left\{\left(\frac{\phi}{2}\right)^{2(q+1)} \left[ \left(\frac{2}{\phi}\right)\frac{\mathrm{d}}{\mathrm{d}(\phi/2)}\right]^q \left[\frac{\sin^2(\phi/2)}{(\phi/2)^2} \right] \right\},
\end{split}
\nonumber
\end{equation}
which corresponds to Eq.\ (\ref{sincderivs}).
\bibliography{latticeDepthMeasurementInandOutOfWDLStaticLattice}

\begin{thebibliography}{36}%
\makeatletter
\providecommand \@ifxundefined [1]{%
 \@ifx{#1\undefined}
}%
\providecommand \@ifnum [1]{%
 \ifnum #1\expandafter \@firstoftwo
 \else \expandafter \@secondoftwo
 \fi
}%
\providecommand \@ifx [1]{%
 \ifx #1\expandafter \@firstoftwo
 \else \expandafter \@secondoftwo
 \fi
}%
\providecommand \natexlab [1]{#1}%
\providecommand \enquote  [1]{``#1''}%
\providecommand \bibnamefont  [1]{#1}%
\providecommand \bibfnamefont [1]{#1}%
\providecommand \citenamefont [1]{#1}%
\providecommand \href@noop [0]{\@secondoftwo}%
\providecommand \href [0]{\begingroup \@sanitize@url \@href}%
\providecommand \@href[1]{\@@startlink{#1}\@@href}%
\providecommand \@@href[1]{\endgroup#1\@@endlink}%
\providecommand \@sanitize@url [0]{\catcode `\\12\catcode `\$12\catcode
  `\&12\catcode `\#12\catcode `\^12\catcode `\_12\catcode `\%12\relax}%
\providecommand \@@startlink[1]{}%
\providecommand \@@endlink[0]{}%
\providecommand \url  [0]{\begingroup\@sanitize@url \@url }%
\providecommand \@url [1]{\endgroup\@href {#1}{\urlprefix }}%
\providecommand \urlprefix  [0]{URL }%
\providecommand \Eprint [0]{\href }%
\providecommand \doibase [0]{http://dx.doi.org/}%
\providecommand \selectlanguage [0]{\@gobble}%
\providecommand \bibinfo  [0]{\@secondoftwo}%
\providecommand \bibfield  [0]{\@secondoftwo}%
\providecommand \translation [1]{[#1]}%
\providecommand \BibitemOpen [0]{}%
\providecommand \bibitemStop [0]{}%
\providecommand \bibitemNoStop [0]{.\EOS\space}%
\providecommand \EOS [0]{\spacefactor3000\relax}%
\providecommand \BibitemShut  [1]{\csname bibitem#1\endcsname}%
\let\auto@bib@innerbib\@empty
\bibitem [{\citenamefont {Morsch}\ and\ \citenamefont
  {Oberthaler}(2006)}]{bec_in_optical_lattice_morsch_2006}%
  \BibitemOpen
  \bibfield  {author} {\bibinfo {author} {\bibfnamefont {Oliver}\ \bibnamefont
  {Morsch}}\ and\ \bibinfo {author} {\bibfnamefont {Markus}\ \bibnamefont
  {Oberthaler}},\ }\bibfield  {title} {\enquote {\bibinfo {title} {Dynamics of
  {B}ose-{E}instein condensates in optical lattices},}\ }\href {\doibase
  10.1103/RevModPhys.78.179} {\bibfield  {journal} {\bibinfo  {journal} {Rev.
  Mod. Phys.}\ }\textbf {\bibinfo {volume} {78}},\ \bibinfo {pages} {179}
  (\bibinfo {year} {2006})}\BibitemShut {NoStop}%
\bibitem [{\citenamefont {Mitroy}\ \emph {et~al.}(2010)\citenamefont {Mitroy},
  \citenamefont {Safronova},\ and\ \citenamefont
  {Clark}}]{mitroy_safronova_matrix_elements_2010}%
  \BibitemOpen
  \bibfield  {author} {\bibinfo {author} {\bibfnamefont {J.}~\bibnamefont
  {Mitroy}}, \bibinfo {author} {\bibfnamefont {M.~S.}\ \bibnamefont
  {Safronova}}, \ and\ \bibinfo {author} {\bibfnamefont {Charles~W.}\
  \bibnamefont {Clark}},\ }\bibfield  {title} {\enquote {\bibinfo {title}
  {Theory and applications of atomic and ionic polarizabilities},}\ }\href
  {http://stacks.iop.org/0953-4075/43/i=20/a=202001} {\bibfield  {journal}
  {\bibinfo  {journal} {J. Phys. B: At. Mol. Opt. Phys.}\ }\textbf {\bibinfo
  {volume} {43}},\ \bibinfo {pages} {202001} (\bibinfo {year}
  {2010})}\BibitemShut {NoStop}%
\bibitem [{\citenamefont {Arora}\ \emph {et~al.}(2011)\citenamefont {Arora},
  \citenamefont {Safronova},\ and\ \citenamefont
  {Clark}}]{arora_tuneout_wavelengths_2011}%
  \BibitemOpen
  \bibfield  {author} {\bibinfo {author} {\bibfnamefont {Bindiya}\ \bibnamefont
  {Arora}}, \bibinfo {author} {\bibfnamefont {M.~S.}\ \bibnamefont
  {Safronova}}, \ and\ \bibinfo {author} {\bibfnamefont {Charles~W.}\
  \bibnamefont {Clark}},\ }\bibfield  {title} {\enquote {\bibinfo {title}
  {Tune-out wavelengths of alkali-metal atoms and their applications},}\ }\href
  {\doibase 10.1103/PhysRevA.84.043401} {\bibfield  {journal} {\bibinfo
  {journal} {Phys. Rev. A}\ }\textbf {\bibinfo {volume} {84}},\ \bibinfo
  {pages} {043401} (\bibinfo {year} {2011})}\BibitemShut {NoStop}%
\bibitem [{\citenamefont {Henson}\ \emph {et~al.}(2015)\citenamefont {Henson},
  \citenamefont {Khakimov}, \citenamefont {Dall}, \citenamefont {Baldwin},
  \citenamefont {Tang},\ and\ \citenamefont
  {Truscott}}]{henson_tuneout_helium_2015}%
  \BibitemOpen
  \bibfield  {author} {\bibinfo {author} {\bibfnamefont {B.~M.}\ \bibnamefont
  {Henson}}, \bibinfo {author} {\bibfnamefont {R.~I.}\ \bibnamefont
  {Khakimov}}, \bibinfo {author} {\bibfnamefont {R.~G.}\ \bibnamefont {Dall}},
  \bibinfo {author} {\bibfnamefont {K.~G.~H.}\ \bibnamefont {Baldwin}},
  \bibinfo {author} {\bibfnamefont {Li-Yan}\ \bibnamefont {Tang}}, \ and\
  \bibinfo {author} {\bibfnamefont {A.~G.}\ \bibnamefont {Truscott}},\
  }\bibfield  {title} {\enquote {\bibinfo {title} {Precision measurement for
  metastable helium atoms of the 413 nm tune-out wavelength at which the atomic
  polarizability vanishes},}\ }\href {\doibase 10.1103/PhysRevLett.115.043004}
  {\bibfield  {journal} {\bibinfo  {journal} {Phys. Rev. Lett.}\ }\textbf
  {\bibinfo {volume} {115}},\ \bibinfo {pages} {043004} (\bibinfo {year}
  {2015})}\BibitemShut {NoStop}%
\bibitem [{\citenamefont {Leonard}\ \emph {et~al.}(2015)\citenamefont
  {Leonard}, \citenamefont {Fallon}, \citenamefont {Sackett},\ and\
  \citenamefont {Safronova}}]{safronova_rb_matrix_elements_2015}%
  \BibitemOpen
  \bibfield  {author} {\bibinfo {author} {\bibfnamefont {R.~H.}\ \bibnamefont
  {Leonard}}, \bibinfo {author} {\bibfnamefont {A.~J.}\ \bibnamefont {Fallon}},
  \bibinfo {author} {\bibfnamefont {C.~A.}\ \bibnamefont {Sackett}}, \ and\
  \bibinfo {author} {\bibfnamefont {M.~S.}\ \bibnamefont {Safronova}},\
  }\bibfield  {title} {\enquote {\bibinfo {title} {High-precision measurements
  of the $^{87}\mathrm{Rb}$ {D}-line tune-out wavelength},}\ }\href {\doibase
  10.1103/PhysRevA.92.052501} {\bibfield  {journal} {\bibinfo  {journal} {Phys.
  Rev. A}\ }\textbf {\bibinfo {volume} {92}},\ \bibinfo {pages} {052501}
  (\bibinfo {year} {2015})}\BibitemShut {NoStop}%
\bibitem [{\citenamefont {Clark}\ \emph {et~al.}(2015)\citenamefont {Clark},
  \citenamefont {Ha}, \citenamefont {Xu},\ and\ \citenamefont
  {Chin}}]{clark_bec_matrix_elements_2015}%
  \BibitemOpen
  \bibfield  {author} {\bibinfo {author} {\bibfnamefont {Logan~W.}\
  \bibnamefont {Clark}}, \bibinfo {author} {\bibfnamefont {Li-Chung}\
  \bibnamefont {Ha}}, \bibinfo {author} {\bibfnamefont {Chen-Yu}\ \bibnamefont
  {Xu}}, \ and\ \bibinfo {author} {\bibfnamefont {Cheng}\ \bibnamefont
  {Chin}},\ }\bibfield  {title} {\enquote {\bibinfo {title} {{Quantum Dynamics
  with Spatiotemporal Control of Interactions in a Stable Bose-Einstein
  Condensate}},}\ }\href {\doibase 10.1103/PhysRevLett.115.155301} {\bibfield
  {journal} {\bibinfo  {journal} {Phys. Rev. Lett.}\ }\textbf {\bibinfo
  {volume} {115}},\ \bibinfo {pages} {155301} (\bibinfo {year}
  {2015})}\BibitemShut {NoStop}%
\bibitem [{\citenamefont {Safronova}\ \emph {et~al.}(2011)\citenamefont
  {Safronova}, \citenamefont {Kozlov},\ and\ \citenamefont
  {Clark}}]{safronova_bbr_2011}%
  \BibitemOpen
  \bibfield  {author} {\bibinfo {author} {\bibfnamefont {M.~S.}\ \bibnamefont
  {Safronova}}, \bibinfo {author} {\bibfnamefont {M.~G.}\ \bibnamefont
  {Kozlov}}, \ and\ \bibinfo {author} {\bibfnamefont {Charles~W.}\ \bibnamefont
  {Clark}},\ }\bibfield  {title} {\enquote {\bibinfo {title} {Precision
  calculation of blackbody radiation shifts for optical frequency metrology},}\
  }\href {\doibase 10.1103/PhysRevLett.107.143006} {\bibfield  {journal}
  {\bibinfo  {journal} {Phys. Rev. Lett.}\ }\textbf {\bibinfo {volume} {107}},\
  \bibinfo {pages} {143006} (\bibinfo {year} {2011})}\BibitemShut {NoStop}%
\bibitem [{\citenamefont {Sherman}\ \emph {et~al.}(2012)\citenamefont
  {Sherman}, \citenamefont {Lemke}, \citenamefont {Hinkley}, \citenamefont
  {Pizzocaro}, \citenamefont {Fox}, \citenamefont {Ludlow},\ and\ \citenamefont
  {Oates}}]{sherman_polarizability_lattice_clock_2012}%
  \BibitemOpen
  \bibfield  {author} {\bibinfo {author} {\bibfnamefont {J.~A.}\ \bibnamefont
  {Sherman}}, \bibinfo {author} {\bibfnamefont {N.~D.}\ \bibnamefont {Lemke}},
  \bibinfo {author} {\bibfnamefont {N.}~\bibnamefont {Hinkley}}, \bibinfo
  {author} {\bibfnamefont {M.}~\bibnamefont {Pizzocaro}}, \bibinfo {author}
  {\bibfnamefont {R.~W.}\ \bibnamefont {Fox}}, \bibinfo {author} {\bibfnamefont
  {A.~D.}\ \bibnamefont {Ludlow}}, \ and\ \bibinfo {author} {\bibfnamefont
  {C.~W.}\ \bibnamefont {Oates}},\ }\bibfield  {title} {\enquote {\bibinfo
  {title} {High-accuracy measurement of atomic polarizability in an optical
  lattice clock},}\ }\href {\doibase 10.1103/PhysRevLett.108.153002} {\bibfield
   {journal} {\bibinfo  {journal} {Phys. Rev. Lett.}\ }\textbf {\bibinfo
  {volume} {108}},\ \bibinfo {pages} {153002} (\bibinfo {year}
  {2012})}\BibitemShut {NoStop}%
\bibitem [{\citenamefont {Whiting}\ \emph {et~al.}(2016)\citenamefont
  {Whiting}, \citenamefont {Keaveney}, \citenamefont {Adams},\ and\
  \citenamefont {Hughes}}]{whiting_et_al_matrix_elements_2016}%
  \BibitemOpen
  \bibfield  {author} {\bibinfo {author} {\bibfnamefont {Daniel~J.}\
  \bibnamefont {Whiting}}, \bibinfo {author} {\bibfnamefont {James}\
  \bibnamefont {Keaveney}}, \bibinfo {author} {\bibfnamefont {Charles~S.}\
  \bibnamefont {Adams}}, \ and\ \bibinfo {author} {\bibfnamefont {Ifan~G.}\
  \bibnamefont {Hughes}},\ }\bibfield  {title} {\enquote {\bibinfo {title}
  {{Direct measurement of excited-state dipole matrix elements using
  electromagnetically induced transparency in the hyperfine Paschen-Back
  regime}},}\ }\href {\doibase 10.1103/PhysRevA.93.043854} {\bibfield
  {journal} {\bibinfo  {journal} {Phys. Rev. A}\ }\textbf {\bibinfo {volume}
  {93}},\ \bibinfo {pages} {043854} (\bibinfo {year} {2016})}\BibitemShut
  {NoStop}%
\bibitem [{\citenamefont {Cronin}\ \emph {et~al.}(2009)\citenamefont {Cronin},
  \citenamefont {Schmiedmayer},\ and\ \citenamefont
  {Pritchard}}]{Cronin_2009_atom_interferometry}%
  \BibitemOpen
  \bibfield  {author} {\bibinfo {author} {\bibfnamefont {Alexander~D.}\
  \bibnamefont {Cronin}}, \bibinfo {author} {\bibfnamefont {J\"org}\
  \bibnamefont {Schmiedmayer}}, \ and\ \bibinfo {author} {\bibfnamefont
  {David~E.}\ \bibnamefont {Pritchard}},\ }\bibfield  {title} {\enquote
  {\bibinfo {title} {Optics and interferometry with atoms and molecules},}\
  }\href {\doibase 10.1103/RevModPhys.81.1051} {\bibfield  {journal} {\bibinfo
  {journal} {Rev. Mod. Phys.}\ }\textbf {\bibinfo {volume} {81}},\ \bibinfo
  {pages} {1051} (\bibinfo {year} {2009})}\BibitemShut {NoStop}%
\bibitem [{\citenamefont {{A. Miffre and M. Jacquey and M. B\"{u}chner and G.
  Tr\'{e}nec and J. Vigu\'{e}}}(2006)}]{atom_interferometry_2006}%
  \BibitemOpen
  \bibfield  {author} {\bibinfo {author} {\bibnamefont {{A. Miffre and M.
  Jacquey and M. B\"{u}chner and G. Tr\'{e}nec and J. Vigu\'{e}}}},\ }\bibfield
   {title} {\enquote {\bibinfo {title} {{Atom interferometry}},}\ }\href
  {\doibase 10.1088/0031-8949/74/2/N01} {\bibfield  {journal} {\bibinfo
  {journal} {Physica Scripta}\ }\textbf {\bibinfo {volume} {74}},\ \bibinfo
  {pages} {C15} (\bibinfo {year} {2006})}\BibitemShut {NoStop}%
\bibitem [{\citenamefont {Bloch}\ \emph {et~al.}(2008)\citenamefont {Bloch},
  \citenamefont {Dalibard},\ and\ \citenamefont
  {Zwerger}}]{bloch_many_body_physics_lattices_2008}%
  \BibitemOpen
  \bibfield  {author} {\bibinfo {author} {\bibfnamefont {Immanuel}\
  \bibnamefont {Bloch}}, \bibinfo {author} {\bibfnamefont {Jean}\ \bibnamefont
  {Dalibard}}, \ and\ \bibinfo {author} {\bibfnamefont {Wilhelm}\ \bibnamefont
  {Zwerger}},\ }\bibfield  {title} {\enquote {\bibinfo {title} {Many-body
  physics with ultracold gases},}\ }\href {\doibase 10.1103/RevModPhys.80.885}
  {\bibfield  {journal} {\bibinfo  {journal} {Rev. Mod. Phys.}\ }\textbf
  {\bibinfo {volume} {80}},\ \bibinfo {pages} {885} (\bibinfo {year}
  {2008})}\BibitemShut {NoStop}%
\bibitem [{\citenamefont {Jo}\ \emph {et~al.}(2012)\citenamefont {Jo},
  \citenamefont {Guzman}, \citenamefont {Thomas}, \citenamefont {Hosur},
  \citenamefont {Vishwanath},\ and\ \citenamefont
  {Stamper-Kurn}}]{gyuboong_kagome_lattice_2012}%
  \BibitemOpen
  \bibfield  {author} {\bibinfo {author} {\bibfnamefont {Gyu-Boong}\
  \bibnamefont {Jo}}, \bibinfo {author} {\bibfnamefont {Jennie}\ \bibnamefont
  {Guzman}}, \bibinfo {author} {\bibfnamefont {Claire~K.}\ \bibnamefont
  {Thomas}}, \bibinfo {author} {\bibfnamefont {Pavan}\ \bibnamefont {Hosur}},
  \bibinfo {author} {\bibfnamefont {Ashvin}\ \bibnamefont {Vishwanath}}, \ and\
  \bibinfo {author} {\bibfnamefont {Dan~M.}\ \bibnamefont {Stamper-Kurn}},\
  }\bibfield  {title} {\enquote {\bibinfo {title} {{Ultracold Atoms in a
  Tunable Optical Kagome Lattice}},}\ }\href {\doibase
  10.1103/PhysRevLett.108.045305} {\bibfield  {journal} {\bibinfo  {journal}
  {Phys. Rev. Lett.}\ }\textbf {\bibinfo {volume} {108}},\ \bibinfo {pages}
  {045305} (\bibinfo {year} {2012})}\BibitemShut {NoStop}%
\bibitem [{\citenamefont {Cahn}\ \emph {et~al.}(1997)\citenamefont {Cahn},
  \citenamefont {Kumarakrishnan}, \citenamefont {Shim}, \citenamefont
  {Sleator}, \citenamefont {Berman},\ and\ \citenamefont
  {Dubetsky}}]{cahn_kaptizadirac_1997}%
  \BibitemOpen
  \bibfield  {author} {\bibinfo {author} {\bibfnamefont {S.~B.}\ \bibnamefont
  {Cahn}}, \bibinfo {author} {\bibfnamefont {A.}~\bibnamefont
  {Kumarakrishnan}}, \bibinfo {author} {\bibfnamefont {U.}~\bibnamefont
  {Shim}}, \bibinfo {author} {\bibfnamefont {T.}~\bibnamefont {Sleator}},
  \bibinfo {author} {\bibfnamefont {P.~R.}\ \bibnamefont {Berman}}, \ and\
  \bibinfo {author} {\bibfnamefont {B.}~\bibnamefont {Dubetsky}},\ }\bibfield
  {title} {\enquote {\bibinfo {title} {Time-domain de {B}roglie wave
  interferometry},}\ }\href {\doibase 10.1103/PhysRevLett.79.784} {\bibfield
  {journal} {\bibinfo  {journal} {Phys. Rev. Lett.}\ }\textbf {\bibinfo
  {volume} {79}},\ \bibinfo {pages} {784} (\bibinfo {year} {1997})}\BibitemShut
  {NoStop}%
\bibitem [{\citenamefont {Gadway}\ \emph {et~al.}(2009)\citenamefont {Gadway},
  \citenamefont {Pertot}, \citenamefont {Reimann}, \citenamefont {Cohen},\ and\
  \citenamefont {Schneble}}]{gadway_kapitzadirac_2009}%
  \BibitemOpen
  \bibfield  {author} {\bibinfo {author} {\bibfnamefont {Bryce}\ \bibnamefont
  {Gadway}}, \bibinfo {author} {\bibfnamefont {Daniel}\ \bibnamefont {Pertot}},
  \bibinfo {author} {\bibfnamefont {Ren\'{e}}\ \bibnamefont {Reimann}},
  \bibinfo {author} {\bibfnamefont {Martin~G.}\ \bibnamefont {Cohen}}, \ and\
  \bibinfo {author} {\bibfnamefont {Dominik}\ \bibnamefont {Schneble}},\
  }\bibfield  {title} {\enquote {\bibinfo {title} {Analysis of
  {K}apitza-{D}irac diffraction patterns beyond the {R}aman-{N}ath regime},}\
  }\href {\doibase 10.1364/OE.17.019173} {\bibfield  {journal} {\bibinfo
  {journal} {Opt. Express}\ }\textbf {\bibinfo {volume} {17}},\ \bibinfo
  {pages} {19173} (\bibinfo {year} {2009})}\BibitemShut {NoStop}%
\bibitem [{\citenamefont {Birkl}\ \emph {et~al.}(1995)\citenamefont {Birkl},
  \citenamefont {Gatzke}, \citenamefont {Deutsch}, \citenamefont {Rolston},\
  and\ \citenamefont {Phillips}}]{birkl_bragg_scattering_optical_lattice_1995}%
  \BibitemOpen
  \bibfield  {author} {\bibinfo {author} {\bibfnamefont {G.}~\bibnamefont
  {Birkl}}, \bibinfo {author} {\bibfnamefont {M.}~\bibnamefont {Gatzke}},
  \bibinfo {author} {\bibfnamefont {I.~H.}\ \bibnamefont {Deutsch}}, \bibinfo
  {author} {\bibfnamefont {S.~L.}\ \bibnamefont {Rolston}}, \ and\ \bibinfo
  {author} {\bibfnamefont {W.~D.}\ \bibnamefont {Phillips}},\ }\bibfield
  {title} {\enquote {\bibinfo {title} {Bragg scattering from atoms in optical
  lattices},}\ }\href {\doibase 10.1103/PhysRevLett.75.2823} {\bibfield
  {journal} {\bibinfo  {journal} {Phys. Rev. Lett.}\ }\textbf {\bibinfo
  {volume} {75}},\ \bibinfo {pages} {2823} (\bibinfo {year}
  {1995})}\BibitemShut {NoStop}%
\bibitem [{\citenamefont {Cheiney}\ \emph {et~al.}(2013)\citenamefont
  {Cheiney}, \citenamefont {Fabre}, \citenamefont {Vermersch}, \citenamefont
  {Gattobigio}, \citenamefont {Mathevet}, \citenamefont {Lahaye},\ and\
  \citenamefont {Gu\'ery-Odelin}}]{cheiney_matterwave_scattering_2013}%
  \BibitemOpen
  \bibfield  {author} {\bibinfo {author} {\bibfnamefont {P.}~\bibnamefont
  {Cheiney}}, \bibinfo {author} {\bibfnamefont {C.~M.}\ \bibnamefont {Fabre}},
  \bibinfo {author} {\bibfnamefont {F.}~\bibnamefont {Vermersch}}, \bibinfo
  {author} {\bibfnamefont {G.~L.}\ \bibnamefont {Gattobigio}}, \bibinfo
  {author} {\bibfnamefont {R.}~\bibnamefont {Mathevet}}, \bibinfo {author}
  {\bibfnamefont {T.}~\bibnamefont {Lahaye}}, \ and\ \bibinfo {author}
  {\bibfnamefont {D.}~\bibnamefont {Gu\'ery-Odelin}},\ }\bibfield  {title}
  {\enquote {\bibinfo {title} {Matter-wave scattering on an amplitude-modulated
  optical lattice},}\ }\href {\doibase 10.1103/PhysRevA.87.013623} {\bibfield
  {journal} {\bibinfo  {journal} {Phys. Rev. A}\ }\textbf {\bibinfo {volume}
  {87}},\ \bibinfo {pages} {013623} (\bibinfo {year} {2013})}\BibitemShut
  {NoStop}%
\bibitem [{\citenamefont {Friebel}\ \emph {et~al.}(1998)\citenamefont
  {Friebel}, \citenamefont {D'Andrea}, \citenamefont {Walz}, \citenamefont
  {Weitz},\ and\ \citenamefont {H\"ansch}}]{friebel_parametric_heating_1998}%
  \BibitemOpen
  \bibfield  {author} {\bibinfo {author} {\bibfnamefont {S.}~\bibnamefont
  {Friebel}}, \bibinfo {author} {\bibfnamefont {C.}~\bibnamefont {D'Andrea}},
  \bibinfo {author} {\bibfnamefont {J.}~\bibnamefont {Walz}}, \bibinfo {author}
  {\bibfnamefont {M.}~\bibnamefont {Weitz}}, \ and\ \bibinfo {author}
  {\bibfnamefont {T.~W.}\ \bibnamefont {H\"ansch}},\ }\bibfield  {title}
  {\enquote {\bibinfo {title} {${\mathrm{co}}_{2}$-laser optical lattice with
  cold rubidium atoms},}\ }\href {\doibase 10.1103/PhysRevA.57.R20} {\bibfield
  {journal} {\bibinfo  {journal} {Phys. Rev. A}\ }\textbf {\bibinfo {volume}
  {57}},\ \bibinfo {pages} {R20--R23} (\bibinfo {year} {1998})}\BibitemShut
  {NoStop}%
\bibitem [{\citenamefont {Ovchinnikov}\ \emph {et~al.}(1999)\citenamefont
  {Ovchinnikov}, \citenamefont {M\"uller}, \citenamefont {Doery}, \citenamefont
  {Vredenbregt}, \citenamefont {Helmerson}, \citenamefont {Rolston},\ and\
  \citenamefont {Phillips}}]{ovchinnikov_rabi_oscillations_1999}%
  \BibitemOpen
  \bibfield  {author} {\bibinfo {author} {\bibfnamefont {Yu.~B.}\ \bibnamefont
  {Ovchinnikov}}, \bibinfo {author} {\bibfnamefont {J.~H.}\ \bibnamefont
  {M\"uller}}, \bibinfo {author} {\bibfnamefont {M.~R.}\ \bibnamefont {Doery}},
  \bibinfo {author} {\bibfnamefont {E.~J.~D.}\ \bibnamefont {Vredenbregt}},
  \bibinfo {author} {\bibfnamefont {K.}~\bibnamefont {Helmerson}}, \bibinfo
  {author} {\bibfnamefont {S.~L.}\ \bibnamefont {Rolston}}, \ and\ \bibinfo
  {author} {\bibfnamefont {W.~D.}\ \bibnamefont {Phillips}},\ }\bibfield
  {title} {\enquote {\bibinfo {title} {{Diffraction of a Released Bose-Einstein
  Condensate by a Pulsed Standing Light Wave}},}\ }\href {\doibase
  10.1103/PhysRevLett.83.284} {\bibfield  {journal} {\bibinfo  {journal} {Phys.
  Rev. Lett.}\ }\textbf {\bibinfo {volume} {83}},\ \bibinfo {pages} {284--287}
  (\bibinfo {year} {1999})}\BibitemShut {NoStop}%
\bibitem [{\citenamefont {Cabrera-Guti\'errez}\ \emph
  {et~al.}(2018)\citenamefont {Cabrera-Guti\'errez}, \citenamefont {Michon},
  \citenamefont {Brunaud}, \citenamefont {Kawalec}, \citenamefont {Fortun},
  \citenamefont {Arnal}, \citenamefont {Billy},\ and\ \citenamefont
  {Gu\'ery-Odelin}}]{cabreragutierez_sudden_phase_shift_2018}%
  \BibitemOpen
  \bibfield  {author} {\bibinfo {author} {\bibfnamefont {C.}~\bibnamefont
  {Cabrera-Guti\'errez}}, \bibinfo {author} {\bibfnamefont {E.}~\bibnamefont
  {Michon}}, \bibinfo {author} {\bibfnamefont {V.}~\bibnamefont {Brunaud}},
  \bibinfo {author} {\bibfnamefont {T.}~\bibnamefont {Kawalec}}, \bibinfo
  {author} {\bibfnamefont {A.}~\bibnamefont {Fortun}}, \bibinfo {author}
  {\bibfnamefont {M.}~\bibnamefont {Arnal}}, \bibinfo {author} {\bibfnamefont
  {J.}~\bibnamefont {Billy}}, \ and\ \bibinfo {author} {\bibfnamefont
  {D.}~\bibnamefont {Gu\'ery-Odelin}},\ }\bibfield  {title} {\enquote {\bibinfo
  {title} {Robust calibration of an optical-lattice depth based on a phase
  shift},}\ }\href {\doibase 10.1103/PhysRevA.97.043617} {\bibfield  {journal}
  {\bibinfo  {journal} {Phys. Rev. A}\ }\textbf {\bibinfo {volume} {97}},\
  \bibinfo {pages} {043617} (\bibinfo {year} {2018})}\BibitemShut {NoStop}%
\bibitem [{\citenamefont {Herold}\ \emph {et~al.}(2012)\citenamefont {Herold},
  \citenamefont {Vaidya}, \citenamefont {Li}, \citenamefont {Rolston},
  \citenamefont {Porto},\ and\ \citenamefont
  {Safronova}}]{herold_matrix_elements}%
  \BibitemOpen
  \bibfield  {author} {\bibinfo {author} {\bibfnamefont {C.~D.}\ \bibnamefont
  {Herold}}, \bibinfo {author} {\bibfnamefont {V.~D.}\ \bibnamefont {Vaidya}},
  \bibinfo {author} {\bibfnamefont {X.}~\bibnamefont {Li}}, \bibinfo {author}
  {\bibfnamefont {S.~L.}\ \bibnamefont {Rolston}}, \bibinfo {author}
  {\bibfnamefont {J.~V.}\ \bibnamefont {Porto}}, \ and\ \bibinfo {author}
  {\bibfnamefont {M.~S.}\ \bibnamefont {Safronova}},\ }\bibfield  {title}
  {\enquote {\bibinfo {title} {{Precision Measurement of Transition Matrix
  Elements via Light Shift Cancellation}},}\ }\href {\doibase
  10.1103/PhysRevLett.109.243003} {\bibfield  {journal} {\bibinfo  {journal}
  {Phys. Rev. Lett.}\ }\textbf {\bibinfo {volume} {109}},\ \bibinfo {pages}
  {243003} (\bibinfo {year} {2012})}\BibitemShut {NoStop}%
\bibitem [{\citenamefont {Kao}\ \emph {et~al.}(2017)\citenamefont {Kao},
  \citenamefont {Tang}, \citenamefont {Burdick},\ and\ \citenamefont
  {Lev}}]{kao_tuneout_dysprosium}%
  \BibitemOpen
  \bibfield  {author} {\bibinfo {author} {\bibfnamefont {Wil}\ \bibnamefont
  {Kao}}, \bibinfo {author} {\bibfnamefont {Yijun}\ \bibnamefont {Tang}},
  \bibinfo {author} {\bibfnamefont {Nathaniel~Q.}\ \bibnamefont {Burdick}}, \
  and\ \bibinfo {author} {\bibfnamefont {Benjamin~L.}\ \bibnamefont {Lev}},\
  }\bibfield  {title} {\enquote {\bibinfo {title} {Anisotropic dependence of
  tune-out wavelength near {Dy} 741-nm transition},}\ }\href {\doibase
  10.1364/OE.25.003411} {\bibfield  {journal} {\bibinfo  {journal} {Opt.
  Express}\ }\textbf {\bibinfo {volume} {25}},\ \bibinfo {pages} {3411}
  (\bibinfo {year} {2017})}\BibitemShut {NoStop}%
\bibitem [{\citenamefont {Beswick}\ \emph {et~al.}(2019)\citenamefont
  {Beswick}, \citenamefont {Hughes},\ and\ \citenamefont
  {Gardiner}}]{beswick_et_al_multipulse}%
  \BibitemOpen
  \bibfield  {author} {\bibinfo {author} {\bibfnamefont {Benjamin~T.}\
  \bibnamefont {Beswick}}, \bibinfo {author} {\bibfnamefont {Ifan~G.}\
  \bibnamefont {Hughes}}, \ and\ \bibinfo {author} {\bibfnamefont {Simon~A.}\
  \bibnamefont {Gardiner}},\ }\bibfield  {title} {\enquote {\bibinfo {title}
  {Lattice-depth measurement using multipulse atom diffraction in and beyond
  the weakly diffracting limit},}\ }\href {\doibase 10.1103/PhysRevA.99.013614}
  {\bibfield  {journal} {\bibinfo  {journal} {Phys. Rev. A}\ }\textbf {\bibinfo
  {volume} {99}},\ \bibinfo {pages} {013614} (\bibinfo {year}
  {2019})}\BibitemShut {NoStop}%
\bibitem [{\citenamefont {Bienert}\ \emph {et~al.}(2003)\citenamefont
  {Bienert}, \citenamefont {Haug}, \citenamefont {Schleich},\ and\
  \citenamefont {Raizen}}]{kicked_rotor_wigner}%
  \BibitemOpen
  \bibfield  {author} {\bibinfo {author} {\bibfnamefont {M.}~\bibnamefont
  {Bienert}}, \bibinfo {author} {\bibfnamefont {F.}~\bibnamefont {Haug}},
  \bibinfo {author} {\bibfnamefont {W.~P.}\ \bibnamefont {Schleich}}, \ and\
  \bibinfo {author} {\bibfnamefont {M.~G.}\ \bibnamefont {Raizen}},\ }\bibfield
   {title} {\enquote {\bibinfo {title} {{Kicked rotor in Wigner phase
  space}},}\ }\href {\doibase 10.1002/prop.200310065} {\bibfield  {journal}
  {\bibinfo  {journal} {Fortschr. Phys.}\ }\textbf {\bibinfo {volume} {51}},\
  \bibinfo {pages} {No. 4–5, 474 – 486} (\bibinfo {year}
  {2003})}\BibitemShut {NoStop}%
\bibitem [{\citenamefont {Beswick}\ \emph {et~al.}(2016)\citenamefont
  {Beswick}, \citenamefont {Hughes}, \citenamefont {Gardiner}, \citenamefont
  {Astier}, \citenamefont {Andersen},\ and\ \citenamefont
  {Daszuta}}]{beswick_et_al}%
  \BibitemOpen
  \bibfield  {author} {\bibinfo {author} {\bibfnamefont {Benjamin~T.}\
  \bibnamefont {Beswick}}, \bibinfo {author} {\bibfnamefont {Ifan~G.}\
  \bibnamefont {Hughes}}, \bibinfo {author} {\bibfnamefont {Simon~A.}\
  \bibnamefont {Gardiner}}, \bibinfo {author} {\bibfnamefont {Hippolyte P.
  A.~G.}\ \bibnamefont {Astier}}, \bibinfo {author} {\bibfnamefont {Mikkel~F.}\
  \bibnamefont {Andersen}}, \ and\ \bibinfo {author} {\bibfnamefont {Boris}\
  \bibnamefont {Daszuta}},\ }\bibfield  {title} {\enquote {\bibinfo {title}
  {$\ensuremath{\epsilon}$-pseudoclassical model for quantum resonances in a
  cold dilute atomic gas periodically driven by finite-duration standing-wave
  laser pulses},}\ }\href {\doibase 10.1103/PhysRevA.94.063604} {\bibfield
  {journal} {\bibinfo  {journal} {Phys. Rev. A}\ }\textbf {\bibinfo {volume}
  {94}},\ \bibinfo {pages} {063604} (\bibinfo {year} {2016})}\BibitemShut
  {NoStop}%
\bibitem [{\citenamefont {Ashcroft}\ and\ \citenamefont
  {Mermin}(1976)}]{ashcroft_mermin}%
  \BibitemOpen
  \bibfield  {author} {\bibinfo {author} {\bibfnamefont {N.W.}\ \bibnamefont
  {Ashcroft}}\ and\ \bibinfo {author} {\bibfnamefont {N.D.}\ \bibnamefont
  {Mermin}},\ }\href@noop {} {\emph {\bibinfo {title} {{Solid State
  Physics}}}}\ (\bibinfo  {publisher} {Saunders College},\ \bibinfo {address}
  {Philadelphia},\ \bibinfo {year} {1976})\BibitemShut {NoStop}%
\bibitem [{\citenamefont {Bach}\ \emph {et~al.}(2005)\citenamefont {Bach},
  \citenamefont {Burnett}, \citenamefont {d'Arcy},\ and\ \citenamefont
  {Gardiner}}]{bach_burnett_d'arcy_gardiner_2005}%
  \BibitemOpen
  \bibfield  {author} {\bibinfo {author} {\bibfnamefont {R.}~\bibnamefont
  {Bach}}, \bibinfo {author} {\bibfnamefont {K.}~\bibnamefont {Burnett}},
  \bibinfo {author} {\bibfnamefont {M.~B.}\ \bibnamefont {d'Arcy}}, \ and\
  \bibinfo {author} {\bibfnamefont {S.~A.}\ \bibnamefont {Gardiner}},\
  }\bibfield  {title} {\enquote {\bibinfo {title} {{Quantum-mechanical cumulant
  dynamics near stable periodic orbits in phase space: Application to the
  classical-like dynamics of quantum accelerator modes}},}\ }\href {\doibase
  10.1103/PhysRevA.71.033417} {\bibfield  {journal} {\bibinfo  {journal} {Phys.
  Rev. A}\ }\textbf {\bibinfo {volume} {71}},\ \bibinfo {pages} {033417}
  (\bibinfo {year} {2005})}\BibitemShut {NoStop}%
\bibitem [{\citenamefont {Daszuta}\ and\ \citenamefont
  {Andersen}(2012)}]{daszuta_andersen_2012}%
  \BibitemOpen
  \bibfield  {author} {\bibinfo {author} {\bibfnamefont {B.}~\bibnamefont
  {Daszuta}}\ and\ \bibinfo {author} {\bibfnamefont {M.~F.}\ \bibnamefont
  {Andersen}},\ }\bibfield  {title} {\enquote {\bibinfo {title} {{Atom
  interferometry using $\delta$-kicked and finite-duration pulse sequences}},}\
  }\href {\doibase 10.1103/PhysRevA.86.043604} {\bibfield  {journal} {\bibinfo
  {journal} {Phys. Rev. A}\ }\textbf {\bibinfo {volume} {86}},\ \bibinfo
  {pages} {043604} (\bibinfo {year} {2012})}\BibitemShut {NoStop}%
\bibitem [{\citenamefont {Martin}\ \emph {et~al.}(1988)\citenamefont {Martin},
  \citenamefont {Oldaker}, \citenamefont {Miklich},\ and\ \citenamefont
  {Pritchard}}]{martin_bragg_scattering}%
  \BibitemOpen
  \bibfield  {author} {\bibinfo {author} {\bibfnamefont {Peter~J.}\
  \bibnamefont {Martin}}, \bibinfo {author} {\bibfnamefont {Bruce~G.}\
  \bibnamefont {Oldaker}}, \bibinfo {author} {\bibfnamefont {Andrew~H.}\
  \bibnamefont {Miklich}}, \ and\ \bibinfo {author} {\bibfnamefont {David~E.}\
  \bibnamefont {Pritchard}},\ }\bibfield  {title} {\enquote {\bibinfo {title}
  {Bragg scattering of atoms from a standing light wave},}\ }\href {\doibase
  10.1103/PhysRevLett.60.515} {\bibfield  {journal} {\bibinfo  {journal} {Phys.
  Rev. Lett.}\ }\textbf {\bibinfo {volume} {60}},\ \bibinfo {pages} {515--518}
  (\bibinfo {year} {1988})}\BibitemShut {NoStop}%
\bibitem [{\citenamefont {Giltner}\ \emph {et~al.}(1995)\citenamefont
  {Giltner}, \citenamefont {McGowan},\ and\ \citenamefont
  {Lee}}]{giltner_bragg_atom_interferometer}%
  \BibitemOpen
  \bibfield  {author} {\bibinfo {author} {\bibfnamefont {David~M.}\
  \bibnamefont {Giltner}}, \bibinfo {author} {\bibfnamefont {Roger~W.}\
  \bibnamefont {McGowan}}, \ and\ \bibinfo {author} {\bibfnamefont {Siu~Au}\
  \bibnamefont {Lee}},\ }\bibfield  {title} {\enquote {\bibinfo {title} {Atom
  interferometer based on bragg scattering from standing light waves},}\ }\href
  {\doibase 10.1103/PhysRevLett.75.2638} {\bibfield  {journal} {\bibinfo
  {journal} {Phys. Rev. Lett.}\ }\textbf {\bibinfo {volume} {75}},\ \bibinfo
  {pages} {2638--2641} (\bibinfo {year} {1995})}\BibitemShut {NoStop}%
\bibitem [{bor(1997)}]{borde_in_atom_interferometry_1997}%
  \BibitemOpen
  \bibfield  {title} {\enquote {\bibinfo {title} {{Bord{\'e}, C. J., 1997,
  “Matter-wave interferometers: a synthetic approach}},}\ }in\ \href
  {\doibase 10.1016/B978-012092460-8/50000-X} {\emph {\bibinfo {booktitle}
  {Atom Interferometry}}},\ \bibinfo {editor} {edited by\ \bibinfo {editor}
  {\bibfnamefont {Paul~R.}\ \bibnamefont {Berman}}}\ (\bibinfo  {publisher}
  {Academic Press},\ \bibinfo {address} {San Diego},\ \bibinfo {year} {1997})\
  pp.\ \bibinfo {pages} {257--292}\BibitemShut {NoStop}%
\bibitem [{\citenamefont {Kozuma}\ \emph {et~al.}(1999)\citenamefont {Kozuma},
  \citenamefont {Deng}, \citenamefont {Hagley}, \citenamefont {Wen},
  \citenamefont {Lutwak}, \citenamefont {Helmerson}, \citenamefont {Rolston},\
  and\ \citenamefont
  {Phillips}}]{kozuma_splitting_with_bragg_diffraction_1999}%
  \BibitemOpen
  \bibfield  {author} {\bibinfo {author} {\bibfnamefont {M.}~\bibnamefont
  {Kozuma}}, \bibinfo {author} {\bibfnamefont {L.}~\bibnamefont {Deng}},
  \bibinfo {author} {\bibfnamefont {E.~W.}\ \bibnamefont {Hagley}}, \bibinfo
  {author} {\bibfnamefont {J.}~\bibnamefont {Wen}}, \bibinfo {author}
  {\bibfnamefont {R.}~\bibnamefont {Lutwak}}, \bibinfo {author} {\bibfnamefont
  {K.}~\bibnamefont {Helmerson}}, \bibinfo {author} {\bibfnamefont {S.~L.}\
  \bibnamefont {Rolston}}, \ and\ \bibinfo {author} {\bibfnamefont {W.~D.}\
  \bibnamefont {Phillips}},\ }\bibfield  {title} {\enquote {\bibinfo {title}
  {{Coherent Splitting of Bose-Einstein Condensed Atoms with Optically Induced
  Bragg Diffraction}},}\ }\href {\doibase 10.1103/PhysRevLett.82.871}
  {\bibfield  {journal} {\bibinfo  {journal} {Phys. Rev. Lett.}\ }\textbf
  {\bibinfo {volume} {82}},\ \bibinfo {pages} {871--875} (\bibinfo {year}
  {1999})}\BibitemShut {NoStop}%
\bibitem [{\citenamefont {Gupta}\ \emph {et~al.}(2001)\citenamefont {Gupta},
  \citenamefont {Leanhardt}, \citenamefont {Cronin},\ and\ \citenamefont
  {Pritchard}}]{gupta_coherent_manipulation_standing_light_wave_2001}%
  \BibitemOpen
  \bibfield  {author} {\bibinfo {author} {\bibfnamefont {Subhadeep}\
  \bibnamefont {Gupta}}, \bibinfo {author} {\bibfnamefont {{Aaron E.}}\
  \bibnamefont {Leanhardt}}, \bibinfo {author} {\bibfnamefont {{Alexander D}}\
  \bibnamefont {Cronin}}, \ and\ \bibinfo {author} {\bibfnamefont {{David E.}}\
  \bibnamefont {Pritchard}},\ }\bibfield  {title} {\enquote {\bibinfo {title}
  {Coherent manipulation of atoms with standing light waves},}\ }\href
  {http://www.sciencedirect.com/science/article/pii/S1296214701011799}
  {\bibfield  {journal} {\bibinfo  {journal} {Comptes Rendus de l'Academie des
  Sciences - Series IV: Physics, Astrophysics}\ }\textbf {\bibinfo {volume}
  {2}},\ \bibinfo {pages} {479--495} (\bibinfo {year} {2001})}\BibitemShut
  {NoStop}%
\bibitem [{\citenamefont {Barnett}\ and\ \citenamefont
  {Radmore}(1997)}]{barnett_radmore_methods_qo_1997}%
  \BibitemOpen
  \bibfield  {author} {\bibinfo {author} {\bibfnamefont {S.~M.}\ \bibnamefont
  {Barnett}}\ and\ \bibinfo {author} {\bibfnamefont {P.~M.}\ \bibnamefont
  {Radmore}},\ }\href@noop {} {\emph {\bibinfo {title} {{Methods in Theoretical
  Quantum Optics}}}}\ (\bibinfo  {publisher} {Clarendon Press},\ \bibinfo
  {address} {Oxford},\ \bibinfo {year} {1997})\BibitemShut {NoStop}%
\bibitem [{\citenamefont {Saunders}\ \emph {et~al.}(2007)\citenamefont
  {Saunders}, \citenamefont {Halkyard}, \citenamefont {Challis},\ and\
  \citenamefont {Gardiner}}]{saunders_halkyard_challis_gardiner_2007}%
  \BibitemOpen
  \bibfield  {author} {\bibinfo {author} {\bibfnamefont {M.}~\bibnamefont
  {Saunders}}, \bibinfo {author} {\bibfnamefont {P.~L.}\ \bibnamefont
  {Halkyard}}, \bibinfo {author} {\bibfnamefont {K.~J.}\ \bibnamefont
  {Challis}}, \ and\ \bibinfo {author} {\bibfnamefont {S.~A.}\ \bibnamefont
  {Gardiner}},\ }\bibfield  {title} {\enquote {\bibinfo {title} {{Manifestation
  of quantum resonances and antiresonances in a finite-temperature dilute
  atomic gas}},}\ }\href {\doibase 10.1103/PhysRevA.76.043415} {\bibfield
  {journal} {\bibinfo  {journal} {Phys. Rev. A}\ }\textbf {\bibinfo {volume}
  {76}},\ \bibinfo {pages} {043415} (\bibinfo {year} {2007})}\BibitemShut
  {NoStop}%
\bibitem [{\citenamefont {Hughes}\ and\ \citenamefont
  {Hase}(2010)}]{hughes_hase_measurements_uncertanties_2010}%
  \BibitemOpen
  \bibfield  {author} {\bibinfo {author} {\bibfnamefont {I.~G.}\ \bibnamefont
  {Hughes}}\ and\ \bibinfo {author} {\bibfnamefont {T.~P.~A.}\ \bibnamefont
  {Hase}},\ }\href@noop {} {\emph {\bibinfo {title} {{Measurements and their
  Uncertainties}}}}\ (\bibinfo  {publisher} {Oxford University Press},\
  \bibinfo {address} {New York},\ \bibinfo {year} {2010})\BibitemShut {NoStop}%
\end{thebibliography}%


%
\end{document}